\documentstyle[12pt]{article}

\font\twlgot =eufm10 scaled \magstep1
\font\egtgot =eufm8
\font\sevgot =eufm7
\font\twlmsb =msbm10 scaled \magstep1
\font\egtmsb =msbm8
\font\sevmsb =msbm7

\newfam\gotfam
\def\pgot{\fam\gotfam\twlgot}
\textfont\gotfam\twlgot
\scriptfont\gotfam\egtgot
\scriptscriptfont\gotfam\sevgot
\def\got{\protect\pgot}
\newfam\msbfam
\textfont\msbfam\twlmsb
\scriptfont\msbfam\egtmsb
\scriptscriptfont\msbfam\sevmsb
\def\Bbb{\protect\pBbb}
\def\pBbb{\relax\ifmmode\expandafter\Bb\else\typeout{You cann't use
Bbb in text mode}\fi}
\def\Bb #1{{\fam\msbfam\relax#1}}

\newcommand{\gE}{{\got E}}
\newcommand{\gA}{{\got A}}

\newcommand{\gJ}{{\got J}}

\newcommand{\gP}{{\got P}}

\newcommand{\gS}{{\got S}}
\newcommand{\gF}{{\got F}}
\newcommand{\gL}{{\got L}}

\newcommand{\ccG}{{\got g}}

\def\thebibliography#1{\section*{References}\list
  {[\arabic{enumi}]}{\settowidth\labelwidth{#1}\leftmargin\labelwidth
    \advance\leftmargin\labelsep
    \usecounter{enumi}}
    \def\newblock{\hskip .11em plus .33em minus .07em}
    \sloppy\clubpenalty4000\widowpenalty4000
    \sfcode`\.=1000\relax}

\def\op#1{\mathop{\fam0 #1}\limits}

\newcommand{\id}{{\rm Id\,}}

\newcommand{\beq}{\begin{equation}}
\newcommand{\eeq}{\end{equation}}
\newcommand{\ben}{\begin{eqnarray}}
\newcommand{\een}{\end{eqnarray}}
\newcommand{\be}{\begin{eqnarray*}}
\newcommand{\ee}{\end{eqnarray*}}
\newcommand{\bea}{\begin{eqalph}}
\newcommand{\eea}{\end{eqalph}}
\newcommand{\cC}{{\cal C}}
\newcommand{\cA}{{\cal A}}

\newcommand{\cP}{{\cal P}}
\newcommand{\cD}{{\cal D}}

\newcommand{\cL}{{\cal L}}

\newcommand{\cE}{{\cal E}}
\newcommand{\cH}{{\cal H}}
\newcommand{\cF}{{\cal F}}
\newcommand{\cS}{{\cal S}}
\newcommand{\cO}{{\cal O}}
\newcommand{\cN}{{\cal N}}
\newcommand{\cG}{{\cal G}}
\newcommand{\bL}{{\bf L}}

\newcommand{\bE}{{\bf E}}

\newcommand{\al}{\alpha}
\newcommand{\vr}{\varrho}
\newcommand{\bt}{\beta}
\newcommand{\dl}{\delta}
\newcommand{\la}{\lambda}
\newcommand{\La}{\Lambda}
\newcommand{\f}{\phi}
\newcommand{\om}{\omega}
\newcommand{\Om}{\Omega}
\newcommand{\m}{\mu}

\newcommand{\g}{\gamma}
\newcommand{\G}{\Gamma}
\newcommand{\th}{\theta}

\newcommand{\vt}{\vartheta}

\newcommand{\up}{\upsilon}
\newcommand{\lng}{\langle}
\newcommand{\rng}{\rangle}
\newcommand{\di}{{\rm dim\,}}

\newcommand{\im}{{\rm Im\,}}

\newcommand{\si}{\sigma}
\newcommand{\Si}{\Sigma}
\newcommand{\w}{\wedge}
\newcommand{\wt}{\widetilde}
\newcommand{\wh}{\widehat}
\newcommand{\ol}{\overline}
\newcommand{\dr}{\partial}
\newcommand{\ar}{\op\longrightarrow}
\newcommand{\ot}{\otimes}
\newcommand{\ap}{\approx}
\newcommand{\ve}{\varepsilon}

\newcommand{\rL}{{\rm L}}

\newcommand{\lto}{\leftarrow}
\newcommand{\llr}{\op\longleftarrow}

\newcommand{\rdr}{\stackrel{\leftarrow}{\dr}{}}

\newcounter{eqalph}
\newcounter{equationa}
\newcounter{remark}
\newcounter{example}
\newcounter{theorem}
\newcounter{proposition}
\newcounter{lemma}
\newcounter{corollary}
\newcounter{definition}
\def\theremark{\arabic{remark}}
\def\thetheorem{\arabic{theorem}}

\def\thedefinition{\arabic{definition}}

\newenvironment{theo}{\refstepcounter{theorem}
{\bf Theorem \thetheorem.}\it }{}

\newenvironment{defi}{\refstepcounter{definition}
{\bf Axiom \thedefinition.}\it }{}

\newenvironment{eqalph}{\stepcounter{equation}
\setcounter{equationa}{\value{equation}} \setcounter{equation}{0}

\begin{eqnarray}}{\end{eqnarray}
\setcounter{equation}{\value{equationa}}}

\newcommand{\mar}[1]{}

\begin{document}
\hbox{}

{\parindent=0pt

{\large \bf Axiomatic classical (prequantum) field theory. Jet
formalism}
\bigskip

{\sc G. Sardanashvily}

{\sl Department of Theoretical Physics, Moscow State University,
117234 Moscow, Russia}

\bigskip
\bigskip

\begin{small}

{\bf Abstract.}  In contrast with QFT, classical field theory can
be formulated in a strict mathematical way if one defines even
classical fields as sections of smooth fiber bundles. Formalism of
jet manifolds provides the conventional language of dynamic
systems (nonlinear differential equations and operators) on fiber
bundles. Lagrangian theory on fiber bundles is algebraically
formulated in terms of the variational bicomplex of exterior forms
on jet manifolds where the Euler--Lagrange operator is present as
a coboundary operator. This formulation is generalized to
Lagrangian theory of even and odd fields on graded manifolds.
Cohomology of the variational bicomplex provides a solution of the
global inverse problem of the calculus of variations, states the
first variational formula and Noether's first theorem in a very
general setting of supersymmetries depending on higher-order
derivatives of fields. A theorem on the Koszul--Tate complex of
reducible Noether identities and Noether's inverse second theorem
extend an original field theory to prequantum field-antifield BRST
theory. Particular field models, jet techniques and some quantum
outcomes are discussed.

\end{small}

 }

\bigskip
\bigskip
\bigskip

\centerline{\large \bf Contents}
\bigskip

\noindent
 {\bf I. Introduction}
\medskip

\noindent
 {\bf II. ACFT. The general framework}

\noindent
 1. The main postulate,  2. Jet manifolds,
 3. Jets and connections, 4. Lagrangian theory of even fields,
 5. Odd fields, 6. The algebra of even and odd fields,
 7. Lagrangian theory of even and odd fields,
 8. Noether's first theorem in a general setting,
 9. The Koszul--Tate complex of Noether identities,
 10. Noether's inverse second theorem, 11. BRST extended field theory,
 12. Local BRST cohomology.

\medskip

\noindent
 {\bf III. Particular models}

\noindent
 13. Gauge theory of principal connections, 14. Topological
 Chern--Simons  theory, 15. Topological BF theory, 16. SUSY gauge
 theory, 17. Field theory on composite bundles, 18. Symmetry
 breaking and Higgs fields, 19. Dirac spinor fields,
 20. Natural and gauge-natural bundles. 21. Gauge  gravitation theory,
 22. Covariant Hamiltonian field theory, 23.  Time-dependent mechanics,
 24. Jets of submanifolds, 25. Relativistic mechanics, 26. String theory.

\medskip

\noindent
 {\bf IV. Quantum outcomes}

\noindent 27. Quantum master equation, 28. Gauge fixing procedure,
28. Green function identities.

\newpage

\centerline{\large \bf I. Introduction}

\bigskip
\bigskip

Our final purpose is QFT, whose existent mathematical formulation
meets many problems. Note that, from the physical viewpoint, it
seems more reasonable to study dequantization of quantum fields.
However, we start with classical fields. Firstly, a generating
functional of Green functions in perturbative QFT is depends on an
action functional of classical fields. Secondly, it may happen
that there exist non-quantizable classical fields, e.g., a Higgs
field. Thirdly, classical field theory can be formulated in a
strict mathematical way if one defines even classical fields as
sections of smooth fiber bundles (Axiom \ref{a1}). Jet formalism
\cite{book,kol,kras,ten,sau} provides the conventional language of
classical field theory as dynamic theory on fiber bundles
\cite{ald,bau,bry,ded,fat03,garc,book,gold,got91,got98,herm,kras,kru73,book00,palais,pom,tak1}.
We agree to call it axiomatic classical field theory (henceforth
ACFT). Section II gives its brief exposition. In Sections III and
IV, particular field models, different jet techniques and some
quantum outcomes are discussed.

Bearing in mind quantization, we treat ACFT as Lagrangian theory
(Axiom \ref{a3}) (see item 22 for covariant Hamiltonian field
theory). We are not concerned with solutions of field equations,
but develop ACFT as prequantum field theory. Lagrangian theory on
fiber bundles is algebraically formulated in terms of the
variational bicomplex of exterior forms on jet manifolds
\cite{and,ander,bau,jmp,book00,olv,ijmms,tak2,tul}. This
formulation is generalized to Lagrangian theory of even and odd
variables on graded manifolds (Axiom \ref{a4})
\cite{barn,jmp05,jmp05a,cmp04}. Theorem \ref{t1} on cohomology of
the variational bicomplex provides a solution of the global
inverse problem of the calculus of variations (Theorem \ref{c1}),
states the first variational formula (Theorem \ref{t2}) and, as a
consequence, leads to Noether's first theorem in a very general
setting of supersymmetries depending on higher-order derivatives
of fields (Theorem \ref{t3}).

Quantization of a Lagrangian field theory essentially depends on
its degeneracy \cite{bat81,bat,gom}. Its Euler--Lagrange operator
generally obeys Noether identities which need not be independent,
but satisfy first-stage Noether identities, and so on. Theorem
\ref{t4} on the Koszul--Tate complex of reducible Noether
identities, Noether's inverse second theorem (Theorem \ref{t5}),
and Theorem \ref{t6} on solutions of the master equation extend
ACFT to prequantum field-antifield BRST theory
\cite{barn95,barn,cmp06,bran97,fust}. Its Lagrangian depends on
antifields and ghosts, associated to Noether identities and gauge
symmetries of an original Lagrangian, and obeys the classical
master equation. This prequantum BRST theory can be quantized in
the framework of perturbative QFT in functional integral terms
\cite{bat,fust,gom}. A problem is that functional integrals are
not expressed into jets of fields \cite{mccloud}. However, there
is a certain relation between the algebras of jets of classical
fields and the algebras of quantum fields such that, in
particular, any variational symmetry of a classical Lagrangian
yields the identities which Euclidean Green functions of quantum
fields satisfy \cite{ward,ward2}.

\newpage

\centerline{\large \bf II. ACFT. The general framework}

\section{The main postulate}

Generalizing the geometric formulations of classical gauge theory
and gravitation theory in fiber bundle terms, let us postulate the
following.

\begin{defi} \label{a1} \mar{a1}
Even classical fields are sections of smooth fiber bundles.
\end{defi}

By virtue of Axiom \ref{a1}, ACFT is represented as dynamic theory
on fiber bundles and, therefore, is conventionally formulated in
terms of jets of section of these fiber bundles
\cite{bry,book,kras,palais,pom}. Note that we throughout are in
the category of finite-dimensional smooth real manifolds, which
are Hausdorff, second-countable and, consequently, paracompact.
The paracompactness of manifolds is very essential for our
consideration because of the abstract de Rham theorem on the sheaf
cohomology (see item 4). In particular, analytic manifolds are
also treated as the smooth ones since a paracompact analytic
manifold need not admit the partition of unity by analytic
functions.

\section{Jet manifolds}

Given a smooth fiber bundle $Y\to X$, a $k$-order jet $j^k_xs$ at
a point $x\in X$ is defined as an equivalence class of sections
$s$ of $Y\to X$ identified by $k+1$ terms of their Taylor series
at $x$. A key point is that a set $J^kY$ of all $k$-order jets is
a finite-dimensional smooth manifold coordinated by $(x^\la,y^i,
y^i_\la, \ldots, y^i_{\la_k\ldots\la_1})$, where $(x^\la,y^i)$ are
bundle coordinates on $Y\to X$ and $y^i_{\la_r\ldots\la_1}$ are
coordinates of derivatives, i.e., $y^i_{\la_r\ldots\la_1}\circ
s=\dr_{\la_r}\cdots\dr_{\la_1}s(x)$ \cite{book,ten,sau}.
Accordingly, the infinite order jets are defined as equivalence
classes of sections of a fiber bundle $Y\to X$ identified by their
Taylor series. Infinite order jets form a paracompact Fr\'echet
(not smooth) manifold $J^\infty Y$
\cite{ander,book,kras,ten,tak2}. It coincides with the projective
limit of the inverse system of finite order jet manifolds
\mar{5.10}\beq
X\op\longleftarrow Y\op\longleftarrow J^1Y \longleftarrow \cdots
J^{r-1}Y \op\longleftarrow J^rY\longleftarrow\cdots. \label{5.10}
\eeq

The main advantage of jet formalism is that it enables us to deal
with finite-dimensional jet manifolds instead of
infinite-dimensional spaces of fields. In the framework of jet
formalism, a $k$-order differential equation on a fiber bundle
$Y\to X$ is defined as a closed subbundle $\gE$ of the jet bundle
$J^kY\to X$. Its solution is a section $s$ of $Y\to X$ whose jet
prolongation $J^ks$ lives in $\gE$. A necessary condition of the
existence of a solution of a differential equation $\gE$ is so
called formal integrability of $\gE$ \cite{book,kras,pom}. A
$k$-order differential operator on $Y\to X$ is defined as a
morphism of the jet bundle $J^kY\to X$ to some vector bundle $E\to
X$. However, the kernel of a differential operator (e.g., an
Euler--Lagrange operator) need not be a differential equation in a
strict sense.

Note that there are different notions of jets. Jets of sections
are particular jets of maps \cite{kol,rei} (see item 20) and jets
of submanifolds \cite{book,kras} (see item 24). Let us also
mention the jets of modules over a commutative ring
\cite{kras,book00} which are representative objects of
differential operators on modules \cite{akman,grot,kras}. In
particular, given a smooth manifold $X$, jets of a projective
$C^\infty(X)$-module $P$ of finite rank are exactly jets of
sections of the vector bundle over $X$ whose module of sections is
$P$ in accordance with the Serre--Swan theorem. The notion of jets
is extended to modules over graded commutative rings \cite{book05}
and modules over algebras of operadic type \cite{niep}. Jets of
modules over a noncommutative ring however are not defined
\cite{book05,epr03a}. A definition of higher-order differential
operators in noncommutative geometry also meets a problem
\cite{book05,lunts,epr03}.

\section{Jets and connections}

Jet manifolds provides the language of modern differential
geometry. Due to the canonical bundle monomorphism $J^1Y\to
T^*X\ot VY$ over $Y$, any connection $\G$ on a fiber bundle $Y\to
X$ is represented by a global section
\be
\G=dx^\la\ot(\dr_\al +\G^i_\la(x^\m,y^j)\dr_i)= dx^\la\ot(\dr_\al
+y^i_\la\dr_i)\circ \G
\ee
of the jet bundle $J^1Y\to X$ and {\it vice versa}
\cite{book,book00,ten,sau}. Accordingly, we have the $T^*X\ot
VY$-valued first order differential operator
\be
D= dx^\la\ot(y^i_\al-\G^i_\la(x^\m,y^j)\dr_i)
\ee
on $Y$. It is called the covariant differential.

Classical field theory and time-dependent mechanics, developed as
particular field theory on bundles over $X=\Bbb R$ (see item 23),
involve the concept of a connection in many aspects
\cite{book98,book00}. Quantum theory appeals to an algebraic
notion of a connection on modules and sheaves
\cite{book05,kosz60,book00,pref}. Jets of modules underlie the
notion of a connection on modules over commutative rings. This
notion is equivalent to that of a connection on vector bundles
$Y\to X$ in the case of $C^\infty(X)$-modules of their sections.
In contrast with jets, connections on modules over a
noncommutative ring are also well defined \cite{dub,book05,land}.

\section{Lagrangian theory of even fields}

We restrict our consideration to Lagrangian field theory, i.e.,
field equations are Euler--Lagrange equations. Note that, if a
field model is characterized by a nonvariational operator, the
Koszul--Tate complex of its Noether identities can be constructed
\cite{oper}, and this field model can be extended to the BRST one.

\begin{defi} \label{a3} \mar{a3}
ACFT is Lagrangian theory.
\end{defi}

There is the extensive literature on the calculus of variations
and Lagrangian formalism on fiber bundles in terms of jet
manifolds
\cite{ald,bau,ded,garc,book,gold,got91,got98,herm,kru73,tak1,vin}.
We formulate Lagrangian theory of even fields in algebraic terms
of the variational bicomplex
\cite{and,ander,bau,jmp,book00,olv,ijmms,tak2,tul}. Namely, one
associates to a fiber bundle $Y\to X$ the following graded
differential algebra (henceforth GDA) $\cO^*_\infty Y$.

The inverse system (\ref{5.10}) of jet manifolds yields the direct
system
\mar{5.7}\beq
\cO^*X\op\longrightarrow \cO^*Y \op\longrightarrow \cO_1^*Y
\ar\cdots \cO^*_{r-1}Y \op\longrightarrow
 \cO_r^*Y \longrightarrow\cdots  \label{5.7}
\eeq
of GDAs $\cO_r^*Y$ of exterior forms on jet manifolds $J^rY$. Its
direct limit is the above mentioned GDA $\cO_\infty^*Y$ of all
exterior forms on finite order jet manifolds (local forms in the
terminology of \cite{barn95,barn,bran97}). This GDA is locally
generated by horizontal forms $dx^\la$ and contact forms
$\th^i_\La=dy^i_\La -y^i_{\la+\La}dx^\la$, where
$\La=(\la_k...\la_1)$ denotes a symmetric multi-index, and
$\la+\La=(\la\la_k...\la_1)$. There is the canonical decomposition
of $\cO^*_\infty Y$ into the modules $\cO^{k,m}_\infty Y$ of
$k$-contact and $m$-horizontal forms ($m\leq n=\di X$).
Accordingly, the exterior differential on $\cO_\infty^*Y$ falls
into the sum $d=d_V+d_H$ of the vertical differential
$d_V:\cO^{k,*}_\infty Y\to \cO^{k+1,*}_\infty Y$ and the total one
$d_H:\cO^{*,m}_\infty Y\to \cO^{*,m+1}_\infty Y$. One also
introduces the projector $\vr$ on $\cO^{>0,n}_\infty Y$ such that
$\vr\circ d_H=0$ and the variational operator $\dl=\vr\circ d$ on
$\cO^{*,n}_\infty Y$ such that $\dl\circ d_H=0$, $\dl\circ\dl=0$.
All these operators split the GDA $\cO^*_\infty Y$ into the
variational bicomplex. We consider its subcomplexes
\mar{b317,f3}\ben
&& 0\to\Bbb R\to \cO^0_\infty Y\ar^{d_H}\cO^{0,1}_\infty Y\cdots
\op\longrightarrow^{d_H} \cO^{0,n}_\infty Y \op\longrightarrow^\dl
\bE_1 \op\longrightarrow^\dl \bE_2 \ar \cdots, \label{b317}\\
&& 0\to \cO^{1,0}_\infty Y\ar^{d_H}\cO^{1,1}_\infty Y\cdots
\op\longrightarrow^{d_H} \cO^{1,n}_\infty Y \op\longrightarrow^\vr
\bE_1\to 0, \qquad \bE_k=\vr(\cO^{k,n}_\infty Y). \label{f3}
\een
Their elements $L\in \cO^{0,n}_\infty Y$ and $\dl L\in \bE_1$ are
a finite order Lagrangian on a fiber bundle $Y\to X$ and its
Euler--Lagrange operators, respectively.

The algebraic Poincar\'e lemma \cite{olv,tul} states that the
variational bicomplex $\cO^*_\infty Y$ is locally exact. In order
to obtain its cohomology, one therefore can use the abstract de
Rham theorem on sheaf cohomology \cite{hir} and the fact that $Y$
is a strong deformation retract of $J^\infty Y$, i.e., sheaf
cohomology of $J^\infty Y$ equals that of $Y$ \cite{ander,jmp}. A
problem is that the paracompact space $J^\infty Y$ admits the
partition of unity by functions which do not belong to
$\cO^0_\infty Y$. Therefore, one considers the variational
bicomplex ${\cal Q}^*_\infty Y\supset \cO^*_\infty Y$ whose
elements are locally exterior forms on finite order jet manifolds,
and obtains its cohomology \cite{and,tak2}. Afterwards, the $d_H$-
and $\dl$-cohomology of $\cO^*_\infty Y$ is proved to be
isomorphic to that of ${\cal Q}^*_\infty Y$ \cite{lmp,jmp,ijmms}.
In particular, cohomology of the variational complex (\ref{b317})
equals the de Rham cohomology of $Y$, while the complex (\ref{f3})
is exact.

The exactness of the complex (\ref{f3}) at the last term states
the global first variational formula which, firstly, shows that an
Euler--Lagrange operator $\dl L$ is really a variational operator
of the calculus of variations and, secondly, leads to Noether's
first theorem. Cohomology of the variational complex (\ref{b317})
at the term $\cO^{0,n}_\infty Y$ provides a solution of the global
inverse problem of the calculus of variations on fiber bundles. It
is the cohomology of variationally trivial Lagrangians, which are
locally $d_H$-exact. Note that this cohomology has been also
derived from cohomology of variational sequences of finite jet
order \cite{and,kru98,vitolo}, and in a different way in
\cite{vin}.

Noether's first theorem is stated in a general case of variational
symmetries depending on higher-order derivatives of fields.
Noether's second theorem is also formulated in jet terms in a
general setting \cite{jpa05,fat,fulp}. In the case of reducible
degenerate Lagrangian systems, one however meets a problem of
definition of higher-stage Noether identities. This problem is
solved by constructing their Koszul--Tate complex
\cite{jmp05,jmp05a}, but its construction involves odd antifields
and leads to Grassmann-graded extension of original even field
theory.

\section{Odd fields}

The algebraic formulation of Lagrangian theory of even fields in
terms of the variational bicomplex is generalized to odd fields
\cite{barn95,barn,jmp05,jmp05a,cmp04}. Note that odd fields in
ACFT need not satisfy the standard spin-statistic connection.
These are odd bosonic ghosts and antifields, though there exist
odd Klein transformations bringing them into even fields
\cite{bogol}.

In many field models (e.g., SUSY gauge theory), odd fields unlike
even ones have no geometric feature. ACFT overcomes this
inconsistence. There are different geometric descriptions of odd
fields in terms both of supermanifolds \cite{cia95,franc} and
graded manifolds \cite{cari,cari03,mont,mont05,sardijmp}. Note
that graded manifolds \cite{ber,kost77,man} are not supermanifolds
\cite{bart,bart93,bruz99,rog}, though every graded manifold can be
associated to a DeWitt $H^\infty$-supermanifold, and {\it vice
versa} \cite{bart,batch2,dew}. Both graded manifolds and
supermanifolds are described in terms of sheaves of graded
commutative algebras \cite{bart,book00}. However, graded manifolds
are characterized by sheaves on smooth manifolds, while
supermanifolds are constructed by gluing of sheaves on supervector
spaces. Lie supergroups, vector and principal superbundles are
defined both in the category of graded manifolds
\cite{alm,boy,hern,stavr} and that of supermanifolds
\cite{bart,cook,book00,shiggs}. Let us mention a different
definition of a super Lie group as a Harish--Chandra pair of a Lie
group and a super Lie algebra \cite{carm,del}.

In ACFT, odd and even fields are described on the same level due
to an appropriate extension of the GDA $\cO^*_\infty Y$. Since QFT
deals with linear spaces of fields, let a bundle $Y\to X$ of
classical fields be a vector bundle. Then all jet bundles $J^kY\to
X$ are also vector bundles. Let us consider a subalgebra
$P^*_\infty Y\subset \cO^*_\infty Y$ of exterior forms whose
coefficients are polynomial in fiber coordinates $y^i$, $y^i_\La$
on these bundles. In particular, the commutative ring $P^0_\infty
Y$ consists of polynomials of coordinates $y^i$, $y^i_\La$ with
coefficients in the ring $C^\infty(X)$. One can associate to such
a polynomial a section of the symmetric tensor product $\op\vee^m
(J^kY)^*$ of the dual of some jet bundle $J^kY\to X$, and {\it
vice versa}. Moreover, one can show that any element of
$P^*_\infty Y$ is an element of the Chevalley--Eilenberg
differential calculus over  $P^0_\infty Y$. This construction is
extended to the case of odd fields if, given a vector bundle $F\to
X$ and jet bundles $J^kF\to X$, one considers their exterior
products, whose sections form a graded commutative ring (see item
6). The result is a Grassmann-graded GDA
$\cS^*_\infty[F;Y]\supset\cO^*_\infty Y$ which is split into the
variational bicomplex (see item 7) and, thus, describes Lagrangian
theory both of even and odd fields. We therefore postulate the
following.

\begin{defi} \label{a4} \mar{a4}
The algebra of ACFT of even and odd fields is the GDA
$\cS^*_\infty[F;Y]$ introduced below.
\end{defi}

\section{The algebra of even and odd fields}

Treating odd fields on a smooth manifold $X$, we follow the
Serre--Swan theorem generalized to graded manifolds. It states
that, if a Grassmann $C^\infty(X)$-algebra is the exterior algebra
of some projective $C^\infty(X)$-module of finite rank, it is
isomorphic to the algebra of graded functions on a graded manifold
with a body $X$ \cite{jmp05a}. Note that $X$ need not be compact
\cite{book05,ren,serr}. By virtue of the Batchelor theorem
\cite{bart}, any graded manifold with a body $X$ is isomorphic to
a graded manifold $(X,\gA_F)$ with the structure sheaf $\gA_F$ of
germs of sections of the exterior bundle
\be
\w F^*=\Bbb R\op\oplus F^*\op\oplus\op\w^2 F^*\op\oplus\cdots,
\ee
where $F^*$ is the dual of some vector bundle $F\to X$. In field
models, Batchelor's isomorphism is fixed from the beginning. We
call $(X,\gA_F)$ the simple graded manifold modelled over $F$. Its
ring $\cA_F$ of graded functions consists of sections of the
exterior bundle $\w F^*$. Then the Grassmann-graded
Chevalley--Eilenberg differential calculus
\be
0\to \Bbb R\to \cA_F \ar^d \cS^1[F;X]\ar^d\cdots
\cS^k[F;X]\ar^d\cdots
\ee
over $\cA_F$ can be constructed \cite{fuks,book05}. One can think
of its elements as being graded differential forms on $X$. In
particular, there is a monomorphism $\cO^*X\to \cS^*[F;X]$.
Following suit of an even GDA $P^*_\infty Y$, let us consider
simple graded manifolds $(X,\gA_{J^rF})$ modelled over the vector
bundles $J^rF\to X$. We have the direct system of corresponding
GDAs
\be
 \cS^*[F;X]\ar
\cS^*[J^1F;X]\ar\cdots \cS^*[J^rF;X]\ar\cdots,
\ee
whose direct limit $\cS^*_\infty[F;X]$ is the Grassmann
counterpart of an even GDA $P^*_\infty Y$.

The total algebra of even and odd fields is the graded exterior
product
\mar{f5}\beq
\cP^*_\infty[F;Y]=P^*_\infty Y\op\w_{\cO^*X} \cS^*_\infty[F;X]
\label{f5}
\eeq
of the GDAs $P^*_\infty Y$ and $\cS^*_\infty[F;X]$ over their
common subalgebra $\cO^*X$ \cite{jmp05,cmp04}. In particular,
$\cP^0_\infty[F;Y]$ is a graded commutative $C^\infty(X)$-ring
whose even and odd generating elements are sections of $Y\to X$
and $F\to X$, respectively. Let $(x^\la,y^i, y^i_\La)$ be bundle
coordinates on jet bundles $J^kY\to X$ and $(x^\la, c^a, c^a_\La)$
those on $J^rF\to X$. For simplicity, let these symbols also stand
for local sections $s$ of these bundles such that
$s^i_\La(x)=y^i_\La$ and $s^a_\La(x)=c^a_\La$. Then the GDA
$\cP^*_\infty[F;Y]$ (\ref{f5}) is locally generated by elements
$(y^i,y^i_\La,c^a,c^a_\La, dx^\la,dy^i,dy^i_\La, dc^a, dc^a_\La)$.
By analogy with $(y^i,y^i_\La)$, one can think of odd generating
elements $(c^a,c^a_\La)$ as being (local) odd fields and their
jets.

Note that this definition of jets of odd fields differs both from
the above mentioned notion of jets of modules over a graded
commutative ring \cite{book05} and the definition of jets of
graded fiber bundles \cite{hern,mont05}. However, it enables us to
consider even and odd fields on the same level, and reproduces the
heuristic notion of jets of odd ghosts in Lagrangian BRST theory
\cite{barn,bran01}. Moreover, one can say that sections of vector
bundles $Y\to X$ and $F\to X$ seen as generating elements of the
ring $\cP^0_\infty[F;Y]$ are {\it sui generis} prequantum fields.

In a general setting, if $Y\to X$ is not a vector bundle, we
consider graded manifolds $(J^rY,\gA_{F_r})$ whose bodies are jet
manifolds $J^rY$, and $F_r=J^rY\times J^rF$ is the pull-back onto
$J^rY$ of the jet bundle $J^rF\to X$ \cite{jmp05a,cmp06}. As a
result, we obtain the direct system of GDAs
\mar{v6}\beq
\cS^*[Y\times F;Y]\ar \cS^*[F_1;J^1Y]\ar\cdots
\cS^*[F_r;J^rY]\ar\cdots, \label{v6}
\eeq
whose direct limit is the GDA $\cS^*_\infty[F;Y]$ in Axiom
\ref{a4}. It is a differential calculus over the ring
$\cS^0_\infty[F;Y]$ of graded functions. The monomorphisms
$\cO^*_rY\to \cS^*[F_r;J^rY]$ yield a monomorphism of the direct
system (\ref{5.7}) to that (\ref{v6}) and, consequently, the
monomorphism $\cO^*_\infty Y\to \cS^*_\infty[F;Y]$ of their direct
limits. Moreover, $\cS^*_\infty[F;Y]$ is a $\cO^0_\infty
Y$-algebra. It contains the $C^\infty(X)$-subalgebra
$\cP^*_\infty[F;Y]$ if a fiber bundle $Y\to X$ is affine. The
$\cO^0_\infty Y$-algebra $\cS^*_\infty[F;Y]$ is locally generated
by elements $(c^a, c^a_\La,dx^\la,dy^i,dy^i_\La, dc^a, dc^a_\La)$
with coefficient functions depending on coordinates $(x^\la,
y^i,y^i_\La)$. One calls $(y^i,c^a)$ the local basis for the GDA
$\cS^*_\infty[F;Y]$. We further  use the collective symbol $s^A$
for its elements. Accordingly, $s^A_\La$ denote jets of $s^A$,
$\th^A_\La=ds^A_\La- s^A_{\la+\La}dx^\la$ are contact forms, and
$\dr_A^\La$ are graded derivations of the $\Bbb R$-ring
$\cS^0_\infty[F;Y]$ such that $\dr_{A'}^{\La'}\rfloor
ds^A_\La=\dl_{A'}^A\dl_\La^{\La'}$. The symbol
$[A]=[s^A]=[s^A_\La]$ stands for the Grassmann parity.

\section{Lagrangian theory of even and odd fields}

There is the canonical decomposition of the GDA
$\cS^*_\infty[F;Y]$  into modules $\cS^{k,m}_\infty[F;Y]$ of
$k$-contact and $m$-horizontal graded forms. Accordingly, the
graded exterior differential on $\cS^*_\infty[F;Y]$ falls into the
sum $d=d_V+d_H$ of the vertical differential $d_V$ and the total
differential
\be
&& d_H(\f)= dx^\la\w d_\la(\f),  \qquad d_\la = \dr_\la +
\op\sum_{0\leq|\La|} s^A_{\la+\La}\dr_A^\La, \qquad \f\in
\cS^*_\infty[F;Y],\\
&& d_H\circ h_0= h_0\circ d, \qquad h_0: \cS^*_\infty[F;Y]\to
\cS^{0,*}_\infty[F;Y].
\ee
We also have the graded projection endomorphism $\vr$ of
$\cS^{<0,n}_\infty[F;Y]$ such that $\vr\circ d_H=0$ and the graded
variational operator $\dl=\vr\circ d$ such that $\dl\circ d_H=0$,
$\dl\circ\dl=0$. With these operators the GDA $\cS^*_\infty[F;Y]$
is split into the Grassmann-graded variational bicomplex. It
contains the subcomplexes
\mar{g111,2}\ben
&& 0\to \Bbb R\ar \cS^0_\infty[F;Y]\ar^{d_H}\cS^{0,1}_\infty[F;Y]
\cdots \ar^{d_H} \cS^{0,n}_\infty[F;Y]\ar^\dl \bE_1
=\vr(\cS^{1,n}_\infty[F;Y]), \label{g111} \\
&& 0\to \cS^{1,0}_\infty[F;Y]\ar^{d_H} \cS^{1,1}_\infty[F;Y]\cdots
\ar^{d_H}\cS^{1,n}_\infty[F;Y]\ar^\vr \bE_1\to 0. \label{g112}
\een
One can think of their even elements
\mar{0709,'}\ben
&& L=\cL\om\in \cS^{0,n}_\infty[F;Y], \qquad \om=dx^1\w\cdots \w
dx^n,
\label{0709}\\
&& \dl L= \th^A\w \cE_A\om=\op\sum_{0\leq|\La|}
 (-1)^{|\La|}\th^A\w d_\La (\dr^\La_A L)\om\in \bE_1 \label{0709'}
\een
as being a graded Lagrangian and its Euler--Lagrange operator,
respectively.

The algebraic Poincar\'e lemma states that the complexes
(\ref{g111}) and (\ref{g112}) are locally exact at all the terms,
except $\Bbb R$ \cite{barn,drag,cmp04}. Then one can obtain
cohomology of these complexes in the same manner as that of the
complexes (\ref{b317}) and (\ref{f3}) \cite{cmp06,cmp04,epr05}.

\begin{theo} \label{t1} \mar{t1}
Cohomology of the  variational complex (\ref{g111}) equals the de
Rham cohomology $H^*(Y)$ of $Y$. The complex (\ref{g112}) is
exact.
\end{theo}

Cohomology of the complex (\ref{g111}) at the term
$\cS^{1,n}_\infty[F;Y]$ provides the following solution of the
global inverse problem of the calculus of variation for graded
Lagrangians.

\begin{theo} \label{c1} \mar{c1}
A $\dl$-closed (i.e., variationally trivial) graded density reads
$L_0=h_0\psi + d_H\xi$, $\xi\in \cS^{0,n-1}_\infty[F;Y]$, where
$\psi$ is a non-exact $n$-form on $Y$. In particular, a
$\dl$-closed odd density is $d_H$-exact.
\end{theo}

Exactness of the complex (\ref{g112}) at the last term implies
that any Lagrangian $L$ admits the decomposition
\mar{g99}\beq
dL=\dl L - d_H\Xi,
\qquad \Xi\in \cS^{1,n-1}_\infty[F;Y], \label{g99}\\
\eeq
where $L+\Xi$ is a Lepagean equivalent of $L$ \cite{cmp04}. This
decomposition leads to the first variational formula (Theorem
\ref{t2}) and Noether's first theorem (Theorem \ref{t3}).

\section{Noether's first theorem in a general setting}

Infinitesimal supersymmetries of ACFT, described by the GDA
$\cS^*_\infty[F;Y]$, are defined as contact graded derivations of
the $\Bbb R$-ring $\cS^0_\infty[F;Y]$ \cite{jmp05,cmp04}. Its
graded derivation $\vt$ is called contact if the Lie derivative
$\bL_\vt$ of the GDA $\cS^*_\infty[F;Y]$ preserves the ideal of
contact graded forms. Contact graded derivations take the form
\mar{g105}\beq
\vt=\vt_H+\vt_V=\vt^\la d_\la + (\up^A\dr_A +\op\sum_{|\La|>0}
d_\La\up^A\dr_A^\La), \qquad \up^A=\vt^A-s^a_\m\vt^\m,
\label{g105}
\eeq
where $\vt^\la$, $\vt^A$ are local graded functions. They
constitute the most general class of so called generalized
(depending on derivatives) symmetries. Generalized  symmetries of
differential equations \cite{and93,bry,ibr,kras,olv} and
Lagrangian systems \cite{bry,fat,cmp04,olv} have been intensively
investigated. In Lagrangian field theory, generalized symmetries
are exemplified by BRST transformations
\cite{barn,jmp05,bat,cmp04,gom}.

\begin{theo}  \label{t2} \mar{t2}
It follows from the decomposition (\ref{g99}) that the Lie
derivative $\bL_\vt L$ of a Lagrangian $L$ (\ref{0709}) with
respect to an arbitrary supersymmetry $\vt$ (\ref{g105}) fulfills
the first variational formula
\mar{g107}\beq
\bL_\vt L= \vt_V\rfloor\dl L +d_H(h_0(\vt\rfloor \Xi_L)) + d_V
(\vt_H\rfloor\om)\cL. \label{g107}
\eeq
\end{theo}

In particular, let $\vt$ be a vertical supersymmetry treated as an
infinitesimal variation of dynamic variables. Then the first
variational formula (\ref{g107}) shows that the Euler--Lagrange
equations $\dl L=0$ are variational equations.

A supersymmetry $\vt$ (\ref{g105}) is called a variational
symmetry of a Lagrangian $L$ if the Lie derivative $\bL_\vt L$ of
$L$ is $d_H$-exact. One can show that $\vt$ is a variational
symmetry iff its vertical part $\up_V$ (\ref{g105}) is well.
Therefore, we further restrict our consideration to vertical
supersymmetries
\mar{f6}\beq
\vt=(\up^A\dr_A +\op\sum_{|\La|>0} d_\La\up^A\dr_A^\La).
\label{f6}
\eeq
A glance at the expression (\ref{f6}) shows that a vertical
supersymmetry is an infinite jet prolongation of its first summand
$\up=\up^A\dr_A$, called the generalized vector field.
Substituting $\vt$ (\ref{f6}) into the first variational formula
(\ref{g107}), we come to Noether's first theorem.

\begin{theo} \label{t3} \mar{t3}
If $\vt$ (\ref{f6}) is a variational symmetry of a Lagrangian $L$
(\ref{0709}) (i.e., $\bL_\up L=d_H\si$, $\si\in
\cS^{0,n-1}_\infty$), the weak conservation law
\be
0\ap d_H(h_0(\vt\rfloor\Xi_L)-\si)
\ee
the Noether current $\gJ_\vt=h_0(\vt\rfloor\Xi_L)$ holds on the
shell $\dl L=0$.
\end{theo}

A vertical supersymmetry $\vt$ (\ref{f6}) is called nilpotent if
$\bL_\vt(\bL_\vt\f)=0$ for any horizontal graded form $\f\in
\cS^{0,*}_\infty[F;Y]$. An even supersymmetry is never nilpotent.

For the sake of simplicity, the common symbol further stands for a
generalized vector field $\up$, the contact graded derivation
$\vt$ (\ref{f6}) determined by $\up$  and the Lie derivative
$\bL_\vt$. We agree to call all these operators a graded
derivation of the GDA $\cS^*_\infty[F;Y]$.

\section{The Koszul--Tate complex of Noether identities}

As was mentioned above, quantization of a Lagrangian field theory
essentially depends on its degeneracy. The Euler--Lagrange
operator (\ref{0709'}) generally obeys non-trivial Noether
identities, which need not be independent, but satisfy first-stage
Noether identities, and so on. Thus, there is a hierarchy of
reducible Noether identities. Note that any Euler--Lagrange
operator obeys trivial Noether identities which are defined as
boundaries of a certain chain complex \cite{jmp05a,cmp06,oper}. A
problem is that trivial higher-stage Noether identities need not
be boundaries, unless a certain condition holds.

The notion of reducible Noether identities came from that of
reducible constraints. By analogy with constraints, the
Koszul--Tate complex of reducible Noether identities has been
invented under rather restrictive regularity condition that
Noether identities of arbitrary stage can be locally separated
into independent and dependent ones \cite{fisch89,fisch}. This
condition has also come from the case of constraints locally given
by a finite set of functions which the inverse mapping theorem is
applied to. A problem is that, in contrast with constraints,
Noether identities of any stage are differential operators. They
are locally given by a set of functions and their jet
prolongations on an infinite order jet manifold. Since the latter
is a Fr\'echet, but not Banach manifold, the inverse mapping
theorem fails to be valid.

We show that, if non-trivial Noether identities of any stage are
finitely generated and if they obey a certain homology regularity
condition, one can associate to the Euler--Lagrange operator of a
degenerate Lagrangian system the exact Koszul--Tate complex whose
boundary operator provides all the Noether identities (Theorem
\ref{t4}) \cite{jmp05a,cmp06,jdg}. This complex is an extension of
the original GDA $\cS^*_\infty[F;Y]$ by means of antifields whose
spaces are density-dual to the modules of Noether identities.

Let us introduce the following notation. The density dual of a
vector bundle $E\to X$ is $\ol E^*=E^*\ot\op\w^n T^*X$. Given
vector bundles $E\to X$ and $V\to X$, let $\cS^*_\infty[V\times
F;Y\times E]$ be the extension of the GDA $\cS^*_\infty[F;Y]$
whose additional even and odd generators are sections of $E\to X$
and $V\to X$, respectively. We consider its subalgebra
$\cP^*_\infty[V,F;Y,E]$ with coefficients polynomial in these new
generators. Let us also assume that the vertical tangent bundle
$VY$ of $Y$ admits the splitting $VY=Y\times W$, where $W\to X$ is
a vector bundle. In this case, there no fiber bundles under
consideration whose transition functions vanish on the shell $\dl
L=0$. Let $\ol Y^*$ denote the density-dual of $W$ in this
splitting.

Let $L$ be a Lagrangian (\ref{0709}) and $\dl L$ its
Euler--Lagrange operator (\ref{0709'}). In order to describe
Noether identities which $\dl L$ satisfies, let us enlarge the GDA
$\cS^*_\infty[F;Y]$ to the GDA $\cP^*_\infty[\ol Y^*,F;Y,\ol F^*]$
with the local basis $\{s^A, \ol s_A\}$, $[\ol s_A]=([A]+1){\rm
mod}\,2$. Its elements $\ol s_A$ are called antifields of
antifield number Ant$[\ol s_A]= 1$ \cite{barn,gom}. The GDA
$\cP^*_\infty[\ol Y^*,F;Y,\ol F^*]$ is endowed with the nilpotent
right graded derivation $\ol\dl=\rdr^A \cE_A$. We have the chain
complex
\mar{v042}\beq
0\lto \im\ol\dl \llr^{\ol\dl} \cP^{0,n}_\infty[\ol Y^*;F;Y;\ol
F^*]_1 \llr^{\ol\dl} \cP^{0,n}_\infty[\ol Y^*;F;Y;\ol F^*]_2
\label{v042}
\eeq
of graded densities of antifield number $\leq 2$. Its one-cycles
define the above mentioned Noether identities, which are trivial
iff cycles are boundaries. Accordingly, elements of the first
homology $H_1(\ol\dl)$ of the complex (\ref{v042}) correspond to
non-trivial Noether identities modulo the trivial ones
\cite{jmp05a,cmp06,jdg,oper}. We assume that $H_1(\ol \dl)$ is
finitely generated. Namely,  there exists a projective
Grassmann-graded $C^\infty(X)$-module $\cC_{(0)}\subset H_1(\ol
\dl)$ of finite rank with a local basis $\{\Delta_r\}$ such that
any Noether identity is a corollary of the Noether identities
\mar{v64}\beq
\ol\dl\Delta_r= \op\sum_{0\leq|\La|} \Delta_r^{A,\La} d_\La
\cE_A=0. \label{v64}
\eeq

The Noether identities (\ref{v64}) need not be independent, but
obey first-stage Noether identities described as follows. By
virtue of the Serre--Swan theorem, the module $\cC_{(0)}$ is
isomorphic to a module of sections of the product $\ol V^*\times
\ol E^*$, where $\ol V^*$ and $\ol E^*$ are the density-duals of
some vector bundles $V\to X$ and $E\to X$. Let us enlarge the GDA
$\cP^*_\infty[\ol Y^*,F;Y,\ol F^*]$ to the GDA $\cP^*_\infty[\ol
E^*\times \ol Y^*,F;Y,\ol F^*\times\ol V^*]$ possessing the local
basis $\{s^A,\ol s_A, \ol c_r\}$ of Grassmann parity $[\ol
c_r]=([\Delta_r]+1){\rm mod}\,2$ and antifield number ${\rm
Ant}[\ol c_r]=2$. This GDA is provided with the nilpotent right
graded derivation $\dl_0=\ol\dl + \rdr^r\Delta_r$ such that its
nilpotency condition is equivalent to the Noether identities
(\ref{v64}). Then we have the chain complex
\mar{v66}\ben
&&0\lto \im\ol\dl \llr^{\ol\dl} \cP^{0,n}_\infty[\ol Y^*,F;Y,\ol
F^*]_1\llr^{\dl_0}
\cP^{0,n}_\infty[\ol E^*\times \ol Y^*,F;Y,\ol F^*\times\ol V^*]_2 \label{v66}\\
&& \qquad \llr^{\dl_0} \cP^{0,n}_\infty[\ol E^*\times\ol
Y^*,F;Y,\ol F^*\times\ol V^*]_3 \nonumber
\een
of graded densities of antifield number $\leq 3$. It has the
trivial homology $H_0(\ol\dl_0)$ and $H_1(\ol\dl_0)$. The
two-cycles of this complex define the above mentioned first-stage
Noether identities. They are trivial if cycles are boundaries, but
the converse need not be true, unless a certain homology condition
holds \cite{jmp05a,cmp06,jdg,oper}. If the complex (\ref{v66})
obeys this condition, elements of its second homology
$H_2(\ol\dl_0)$ define non-trivial first-stage Noether identities
modulo the trivial ones. Let us assume that $H_2(\dl_0)$ is
finitely generated. Namely, there exists a projective
Grassmann-graded $C^\infty(X)$-module $\cC_{(1)}\subset
H_2(\dl_0)$ of finite rank with a local basis $\{\Delta_{r_1}\}$
such that any first-stage Noether identity is a corollary of the
equalities
\mar{v82}\beq
\op\sum_{0\leq|\La|} \Delta_{r_1}^{r,\La} d_\La \Delta_r +\ol\dl
h_{r_1} =0. \label{v82}
\eeq

The first-stage Noether identities (\ref{v82}) need not be
independent, but satisfy the second-stage ones, and so on.
Iterating the arguments, we come to the following
\cite{jmp05a,cmp06,jdg}.

\begin{theo} \label{t4} \mar{t4} One can associate to a
degenerate $N$-stage reducible Lagrangian system the exact
Koszul--Tate complex (\ref{v94}) with the boundary operator
(\ref{v92}) whose nilpotency property restarts all Noether and
higher-stage Noether identities (\ref{v64}) and (\ref{v93}) if
these identities are finitely generated and iff this complex obeys
the homology regularity condition.
\end{theo}

Namely, there are vector bundles $V_1,\ldots, V_N, E_1, \ldots,
E_N$ over $X$ and the GDA
\mar{v91}\beq
\ol\cP^*_\infty\{N\}=\cP^*_\infty[\ol E^*_N\times\cdots\times\ol
E^*_1\times\ol E^*\times\ol Y^*,F;Y,\ol F^*\times\ol V^*\times\ol
V^*_1\times\cdots\times\ol V_N^*] \label{v91}
\eeq
with a local basis $\{s^A,\ol s_A, \ol c_r, \ol c_{r_1}, \ldots,
\ol c_{r_N}\}$ of antifield number Ant$[\ol c_{r_k}]=k+2$. Let the
indexes $k=-1,0$ further stand for $\ol s_A$ and $\ol c_r$,
respectively. The GDA $\ol\cP^*_\infty\{N\}$ (\ref{v91}) is
provided with the nilpotent right graded derivation  (the
Koszul--Tate differential)
\mar{v92}\ben
&&\dl_N=\rdr^A\cE_A +
\op\sum_{0\leq|\La|}\rdr^r\Delta_r^{A,\La}\ol s_{\La A} +
\op\sum_{1\leq k\leq N}\rdr^{r_k} \Delta_{r_k},
\label{v92}\\
&& \Delta_{r_k}= \op\sum_{0\leq|\La|}
\Delta_{r_k}^{r_{k-1},\La}\ol c_{\La r_{k-1}} + \op\sum_{0\leq
|\Si|, |\Xi|}(h_{r_k}^{(r_{k-2},\Si)(A,\Xi)}\ol c_{\Si r_{k-2}}\ol
s_{\Xi A}+...), \nonumber
\een
of antifield number -1. With $\dl_N$, we have the exact chain
complex
\mar{v94}\ben
&&0\lto \im \ol\dl \llr^{\ol\dl} \cP^{0,n}_\infty[\ol Y^*,F;Y,\ol
F^*]_1\llr^{\dl_0} \ol\cP^{0,n}_\infty\{0\}_2\llr^{\dl_1}
\ol\cP^{0,n}_\infty\{1\}_3\cdots
\label{v94}\\
&& \qquad
 \llr^{\dl_{N-1}} \ol\cP^{0,n}_\infty\{N-1\}_{N+1}
\llr^{\dl_N} \ol\cP^{0,n}_\infty\{N\}_{N+2}\llr^{\dl_N}
\ol\cP^{0,n}_\infty\{N\}_{N+3}, \nonumber
\een
of graded densities of antifield number $\leq N+3$ which is
assumed to satisfy the homology regularity condition. This
condition states that any $\dl_{k<N-1}$-cycle $\f\in
\ol\cP_\infty^{0,n}\{k\}_{k+3}\subset
\ol\cP_\infty^{0,n}\{k+1\}_{k+3}$ is a $\dl_{k+1}$-boundary. The
nilpotency property of the boundary operator $\dl_N$ (\ref{v92})
implies the Noether identities (\ref{v64}) and the  $(k\leq
N)$-stage Noether identities
\mar{v93}\beq
\op\sum_{0\leq|\La|} \Delta_{r_k}^{r_{k-1},\La}d_\La
(\op\sum_{0\leq|\Si|} \Delta_{r_{k-1}}^{r_{k-2},\Si}\ol c_{\Si
r_{k-2}}) +  \ol\dl(\op\sum_{0\leq |\Si|,
|\Xi|}h_{r_k}^{(r_{k-2},\Si)(A,\Xi)}\ol c_{\Si r_{k-2}}\ol s_{\Xi
A})=0. \label{v93}
\eeq

\section{Noether's inverse second theorem}

Noether's second theorem in different variants relates the Noether
and higher-stage Noether identities to the gauge and higher-stage
gauge symmetries of a Lagrangian system
\cite{jpa05,jmp05,fulp,jdg}. However, the notion of a general
gauge symmetry of a Lagrangian system and, consequently, a
formulation of Noether's direct second theorem meet difficulties.
In particular, it may happen that gauge symmetries are not
assembled into an algebra, or they form an algebra on-shell
\cite{fulp02,gom}. At the same time, Noether identities are well
defined (Theorem \ref{t4}). Therefore, one can Noether's inverse
second theorem (Theorem \ref{t5}) in order to obtain gauge
symmetries of a degenerate Lagrangian system. This theorem
associates to the antifield Koszul--Tate complex (\ref{v94}) the
cochain sequence (\ref{w36}) of ghosts, whose ascent operator
(\ref{w108}) provides gauge and higher-stage gauge symmetries of a
Lagrangian field theory.

Given the GDA $\ol\cP^*_\infty\{N\}$ (\ref{v91}), let us consider
the GDA
\mar{w5}\beq
\cP^*_\infty\{N\}=\cP^*_\infty[V_N\times\cdots V_1\times
V,F;Y,E\times E_1\times\cdots \times E_N] \label{w5}
\eeq
possessing the local basis $\{s^A, c^r, c^{r_1}, \ldots,
c^{r_N}\}$ of Grassmann parity $[c^{r_k}]=([\ol c_{r_k}]+1){\rm
mod}\,2$ and antifield number ${\rm Ant}[c^{r_k}]=-(k+1)$. Its
elements $c^{r_k}$, $k\in\Bbb N$, are called the ghosts of ghost
number gh$[c^{r_k}]=k+1$ \cite{barn,gom}.

\begin{theo} \label{t5} \mar{t5}
Given the Koszul--Tate complex (\ref{v94}), the graded commutative
ring $\cP_\infty^0\{N\}$ is split into the cochain sequence
\mar{w36}\beq
0\to \cS^0_\infty[F;Y]\ar^{u_e} \cP^0_\infty\{N\}_1\ar^{u_e}
\cP^0_\infty\{N\}_2\ar^{u_e}\cdots, \label{w36}
\eeq
with the odd ascent operator
\mar{w108}\ben
&& u_e=u + \op\sum_{1\leq k\leq N} u_{(k)}, \label{w108}\\
&& u= u^A\frac{\dr}{\dr s^A}, \qquad u^A =\op\sum_{0\leq|\La|}
c^r_\La\eta(\Delta^A_r)^\La, \label{w33}\\
&& u_{(k)}= u^{r_{k-1}}\frac{\dr}{\dr c^{r_{k-1}}}, \qquad
u^{r_{k-1}}=\op\sum_{0\leq|\La|}
c^{r_k}_\La\eta(\Delta^{r_{k-1}}_{r_k})^\La, \qquad k=1,\ldots,N,
\label{w38}\\
&&\eta (f)^\La = \op\sum_{0\leq|\Si|\leq k-|\La|}(-1)^{|\Si+\La|}
C^{|\Si|}_{|\Si+\La|} d_\Si f^{\Si+\La}, \qquad
C^a_b=\frac{b!}{a!(b-a)!}. \nonumber
\een
\end{theo}

The components $u$ (\ref{w33}), $u_{(k)}$ (\ref{w38}) of the
ascent operator $u_e$ (\ref{w108}) are the above mentioned gauge
and higher-stage gauge symmetries of an original Lagrangian,
respectively. Indeed, let us consider the total GDA
$P^*_\infty\{N\}$ generated by original fields, ghosts and
antifields
\mar{f10}\beq
\{s^A, c^r, c^{r_1}, \ldots, c^{r_N},\ol s_A,\ol c_r, \ol c_{r_1},
\ldots, \ol c_{r_N}\}. \label{f10}
\eeq
It contains subalgebras $\ol\cP^*_\infty\{N\}$ (\ref{v91}) and
$\cP^*_\infty\{N\}$ (\ref{w5}), whose operators $\dl_N$
(\ref{v92}) and $u_e$ (\ref{w108}) are prolonged to
$P^*_\infty\{N\}$. Let us extend an original Lagrangian $L$ to the
Lagrangian
\mar{w8}\beq
L_e=\cL_e\om=L+L_1=L + \op\sum_{0\leq k\leq N}
c^{r_k}\Delta_{r_k}\om=L +\dl_N( \op\sum_{0\leq k\leq N}
c^{r_k}\ol c_{r_k}\om) \label{w8}
\eeq
of zero antifield number. It is readily observed that the
Koszul--Tate differential $\dl_N$ is a variational symmetry of the
Lagrangian $L_e$ (\ref{w8}), i.e., we have the equalities
\mar{w19,20}\ben
&& \frac{\op\dl^\lto (c^r\Delta_r)}{\dl \ol s_A}\cE_A\om
=u^A\cE_A\om=d_H\si_0, \label{w19}\\
&&  [\frac{\op\dl^\lto (c^{r_i}\Delta_{r_i})}{\dl \ol s_A}\cE_A
+\op\sum_{k<i} \frac{\op\dl^\lto (c^{r_i}\Delta_{r_i})}{\dl \ol
c_{r_k}}\Delta_{r_k}]\om= d_H\si_i, \qquad i=1,\ldots,N.
\label{w20}
\een
A glance at the equality (\ref{w19}) shows that the graded
derivation $u$ (\ref{w33}) is a variational symmetry of an
original Lagrangian $L$. Parameterized by ghosts $c^r$, it is a
gauge symmetry of $L$ \cite{jmp05,cmp04}.  The equalities
(\ref{w20}) are brought into the form
\be
\op\sum_{0\leq|\Si|} d_\Si u^{r_{i-1}}\frac{\dr}{\dr
c^{r_{i-1}}_\Si} u^{r_{i-2}} =\ol\dl(\al^{r_{i-2}}), \qquad
\al^{r_{i-2}} = -\op\sum_{0\leq|\Si|}
\eta(h_{r_i}^{(r_{i-2})(A,\Xi)})^\Si d_\Si(c^{r_i} \ol s_{\Xi A}).
\ee
It follows that graded derivations $u_{(k)}$ (\ref{w38}) are the
$k$-stage gauge symmetries of a reducible Lagrangian system
\cite{jpa05,jmp05,jdg}.

We agree to call $u_e$ (\ref{w108}) the total gauge operator. In
contrast with the Koszul--Tate one, this operator need not be
nilpotent. However, one can say that gauge and higher-stage gauge
symmetries of a Lagrangian system form an algebra (resp. an
algebra on the shell) if the total gauge operator $u_e$ can be
extended to a graded derivation $u_E$ of ghost number 1 which is
nilpotent (resp. nilpotent on the shell)
\cite{ijgmmp05,cmp06,epr04}. It reads
\mar{w109}\beq
u_E=u_e+ \xi= u^A\dr_A + \op\sum_{1\leq k\leq N}(u^{r_{k-1}}
+\xi^{r_{k-1}})\dr_{r_{k-1}}+\xi^{r_N}\dr_{r_N}, \label{w109}
\eeq
where the coefficients $\xi^{r_{k-1}}$ are at least quadratic in
ghosts, and $(u_E\circ u_E)(f)$ is zero (resp. $\ol\dl$-exact) on
graded functions $f\in \cP^0_\infty\{N\}$. For instance, the total
gauge operator in irreducible gauge theory is an operator of gauge
transformations whose parameter functions are replaced with the
ghosts. Its nilpotent extension (\ref{w109}) is a familiar BRST
operator \cite{ijgmmp05,cmp04}.

\section{BRST extended field theory}

ACFT extended to ghosts and antifields exemplifies so called
field-antifield Lagrangian systems of the following type
\cite{barn,gom}.

Given a fiber bundle $Z\to X$ and a vector bundle $Z'\to X$, let
us consider a GDA $\cP^*_\infty[\ol Z^*,Z';Z,\ol Z'^*]$ with a
local basis $\{z^a,\ol z_a\}$, where $[\ol z_a]=([z^a]+1){\rm
mod}\,2$. One can think of its elements $z^a$ and $\ol z_a$ as
being fields and antifields, respectively. Its submodule
$\cP^{0,n}_\infty[\ol Z^*,Z';Z,\ol Z'^*]$ of horizontal densities
is provided with the binary operation
\mar{f11}\beq
\{\gL\om,\gL'\om\}=[\frac{\op\dl^\lto \gL}{\dl \ol z_a}\frac{\dl
\gL'}{\dl z^a} + (-1)^{[\gL][\gL']}\frac{\op\dl^\lto \gL'}{\dl \ol
z_a}\frac{\dl \gL}{\dl z^a}]\om, \label{f11}
\eeq
called the antibracket by analogy with that in field-antifield
BRST theory \cite{gom}. One treats this operation as {\it sui
generis} odd Poisson structure \cite{ala,barn98}. Let us associate
to a Lagrangian $\gL\om$ the odd graded derivations
\mar{w37}\beq
\up_\gL=\frac{\op\dl^\lto \gL}{\dl \ol z_a} \frac{\dr}{\dr z^a},
\qquad \ol\up_\gL=\frac{\op\dr^\lto}{\dr \ol z_a}\frac{\dl
\gL}{\dl z^a}. \label{w37}
\eeq
Then the following conditions are equivalent: (i) the graded
derivation $\up_\gL$ (\ref{w37}) is a variational symmetry of a
Lagrangian $\gL\om$,  (ii) so is the graded derivation
$\ol\up_\gL$, (iii) the (classical) master equation
\mar{w44}\beq
\{\gL\om,\gL\om\}=2\frac{\op\dl^\lto \gL}{\dl \ol z_a}\frac{\dl
\gL}{\dl z^a}\om =d_H\si \label{w44}
\eeq
holds. For instance, any variationally trivial Lagrangian
satisfies  the master equation. We say that a solution of the
master equation is not trivial if no graded derivation (\ref{w37})
vanishes.

Let us return to an original Lagrangian system $L$ and its
extension $L_e$ (\ref{w8}) to antifields and ghosts (\ref{f10}),
together with the odd graded derivations (\ref{w37}) which read
\be
\up_e= \frac{\op\dl^\lto \cL_1}{\dl \ol s_A}\frac{\dr}{\dr s^A} +
\op\sum_{0\leq k\leq N} \frac{\op\dl^\lto \cL_1}{\dl \ol
c_{r_k}}\frac{\dr}{\dr c^{r_k}}, \qquad
 \ol\up_e= \frac{\rdr }{\dr
\ol s_A}\frac{\dl\cL_1}{\dl s^A} + [\frac{\rdr }{\dr \ol
s_A}\frac{\dl\cL}{\dl s^A} +\op\sum_{0\leq k\leq N} \frac{\rdr
}{\dr \ol c_{r_k}}\frac{\dl \cL_1}{\dl c^{r_k}}].
\ee
An original Lagrangian $L$ provides a trivial solution of the
master equation. A goal is to extend it to a nontrivial solution
\mar{w61}\beq
L+L_1+L_2+\cdots =L_e+L' \label{w61}
\eeq
of the master equation by means of terms $L_i$ of polynomial
degree $i>1$ in ghosts and zero antifield number. Such an
extension need not exists. However, one can show the following
\cite{cmp06}.

\begin{theo} \label{t6} \mar{t6}
(i) A solution (\ref{w61}) of the master equation exists only if
the graded derivation $u_e$ (\ref{w108}) is extended to a graded
derivation nilpotent on the shell. (ii) If the total gauge
operator $u_e$ (\ref{w108}) admits a nilpotent extension $u_E$
(\ref{w109}) independent of antifields, then the Lagrangian
\mar{w133}\beq
L_E=L_e + \op\sum_{1\leq k\leq N}\xi^{r_{k-1}}\ol c_{r_{k-1}}\om=
L+u_E( \op\sum_{0\leq k\leq N} c^{r_{k-1}}\ol c_{r_{k-1}})\om
+d_H\si \label{w133}
\eeq
satisfies the master equation $\{L_E,L_E\}=0$, and $\up_e=u_E$
called the BRST operator.
\end{theo}

Let ACFT with a Lagrangian $L$ be extended to antifields and
ghosts (\ref{f10}) which come from Theorems \ref{t4} and \ref{t5},
and let its Lagrangian $L$ admit an extension to a solution of the
master equation (\ref{w44}). One can think of this BRST extended
system  as being prequantum field theory, which is quantized both
in the framework of the Batalin--Vilkoviski (BV) quantization
\cite{bat,gom,fust} and in a different way (see Section IV). For
instance, this is the case of Yang--Mills gauge and SUSY gauge
theories (see item 16).

\section{Relative BRST cohomology}

A solution of the classical master equation (\ref{w44}) is not
unique. At least, it is defined up to variationally trivial
Lagrangians. In particular, let a bundle $Y\to X$ of even
classical fields be affine. Its de Rham cohomology equals that of
$X$. Then by virtue of Theorem \ref{c1}, any variationally trivial
Lagrangian reads $L_0=\f + d_H\xi$ where $\f$ is a non-exact
$n$-form on $X$.

Note that the generating functional in perturbative QFT depends on
the action functional and, thus, is defined with accuracy to
variationally trivial Lagrangians, which are $d_H$-exact in QFT on
$X=\Bbb R^n$. This fact motivates us to treat all Lagrangians in
field-antifield BRST theory up to $d_H$-exact ones. They are
defined as elements of the cohomology at the last term of the
cochain complex
\be
0\to \Bbb R\ar P^0_\infty\{N\}\ar^{d_H}P^{0,1}_\infty\{N\} \cdots
\ar^{d_H} P^{0,n}_\infty\{N\}\to 0.
\ee
Equivalently, one considers so called local functionals $\int
Ld^nx$ which, in the case of even fields, are evaluated for the
jet prolongations of sections of $Y\to X$ of compact support
\cite{ala,barn98,barn,bran97}.

In particular, any vertical graded derivation $\vt$ (\ref{f6})
obeys the relation $\vt\circ d_H= d_H\circ \vt$. If $\vt$ is
nilpotent, we therefore have a complex of complexes of horizontal
graded forms $P^{0,*}_\infty\{N\}$ with respect to the nilpotent
operator $\vt$ and the total differential $d_H$. For instance, let
$\vt$ be a BRST operator. Then one studies $d_H$-relative (local
in the terminology of \cite{barn}) and iterated BRST cohomology
\cite{barn95,barn,bran97,lmp,cmp04}. Relative and iterated
cohomology of graded densities coincide with each other. For
instance, a glance at the formula (\ref{w133}) shows that an
original Lagrangian $L$ and its BRST extension $L_E$ are of the
same relative BRST cohomology class.

\bigskip
\bigskip
\bigskip

\centerline{\large \bf III. Particular models}

\section{Gauge theory of principal connections}

Let us consider gauge theory of principal connections on a
principal bundle $P\to X$ with a structure Lie group $G$.
Principal connections are $G$-equivariant connections on $P\to X$
and, therefore, they are represented by sections of the quotient
bundle
\mar{f30}\beq
C=J^1P/G\to X \label{f30}
\eeq
\cite{book,book00}. This is an affine bundle coordinated by
$(x^\la, a^r_\la)$ such that, given a section $A$ of $C\to X$, its
components $A^r_\la=a^r_\la\circ A$ are coefficients of the
familiar local connection form \cite{kob} (i.e., gauge
potentials). Therefore, one calls $C$ (\ref{f30}) the bundle of
principal connections. A key point is that its first order jet
manifold $J^1C$ admits the canonical splitting over $C$ given by
the coordinate expression
\mar{f31}\beq
a_{\la\m}^r = \frac12\cF^r_{\la\m} +  \frac12\cS^r_{\la\m}
=\frac{1}{2}(a_{\la\m}^r + a_{\m\la}^r
 - c_{pq}^r a_\la^p a_\m^q) + \frac{1}{2}
(a_{\la\m}^r - a_{\m\la}^r + c_{pq}^r a_\la^p a_\m^q), \label{f31}
\eeq
where $c_{pq}^r$ are the structure constants of the Lie algebra
$\ccG$ of $G$, and $F^r_{\la\m}=\cF^r_{\la\m}\circ J^1A$ is the
curvature (the strength (\ref{r47'})) of a principal connection
$A$.

There is a unique (Yang--Mills) quadratic gauge invariant
Lagrangian $L_{YM}$ on $J^1C$ which factorizes through the
component $\cF^r_{\la\m}$ of the splitting (\ref{f31}). It obeys
the irreducible Noether identities
\be
 c^r_{ji}a^i_\la\cE_r^\la + d_\la\cE_j^\la=0.
\ee
The corresponding gauge symmetries are $G$-invariant vertical
vector fields on $P$. They are given by sections $\xi=\xi^re_r$ of
the Lie algebra bundle $V_GP=VP/G$, and define vector fields
\mar{f33}\beq
\xi=(-c^r_{ji}\xi^ja^i_\la + \dr_\la\xi^r)\dr^\la_r \label{f33}
\eeq
on the bundle of principal connections $C$ such that
$\bL_{J^1\xi}L_{YM}=0$. As a consequence, the basis
$(a^r_\la,c^r,\ol a^\la_r,\ol c_r)$ for the BRST extended gauge
theory consists of gauge potentials $a^r_\la$, ghosts $c^r$ of
ghost number 1, and antifields $\ol a^\la_r$, $\ol c_r$ of
antifield numbers 1 and 2, respectively. Replacing gauge
parameters $\xi^r$ in $\xi$ (\ref{f33}) with odd ghost $c^r$, we
obtain the total gauge operator $u_e$ (\ref{w108}), whose
nilpotent extension is the well known BRST operator
\be
u_E= (-c^r_{ji}c^ja^i_\la + c^r_\la)\frac{\dr}{\dr a_\la^r}
-\frac12  c^r_{ij}c^ic^j\frac{\dr}{\dr c^r}.
\ee
Hence, the Yang--Mills Lagrangian is extended to a solution of the
master equation
\be
L_E=L_{YM}+ (-c^r_{ij}c^ja^i_\la + c^r_\la)\ol a^\la_r\om -\frac12
c^r_{ij}c^ic^j\ol c_r\om.
\ee

\section{Topological
 Chern--Simons theory}

Vector fields $\xi$ (\ref{f33}) are variational symmetries of the
Lagrangian of topological Chern--Simons theory. One usually
considers Chern--Simons theory whose Lagrangian is the local
Chern-- Simons form derived from the local transgression formula
for the second Chern characteristic form. The global Chern--Simons
Lagrangian is well defined, but depends on a background gauge
potential \cite{bor,bor06,fat05,mpl}.

The fiber bundle $J^1P\to C$ is a trivial $G$-principal bundle
canonically isomorphic to $C\times P\to C$. It admits the
canonical principal connection
\be
\cA =dx^\la\ot(\dr_\la +a_\la^p \ve_p) + da^r_\la\ot\dr^\la_r
\ee
\cite{gar77,book00}. Its curvature defines the canonical
$V_GP$-valued 2-form
\mar{f34}\beq
\gF =(da_\m^r\w dx^\m + \frac{1}{2} c_{pq}^r a_\la^p a_\m^q
dx^\la\w dx^\m)\ot e_r \label{f34}
\eeq
on $C$. Given a section $A$ of $C\to X$, the pull-back
\mar{r47'}\beq
F_A=A^*\gF=\frac12 F^r_{\la\m}dx^\la\w dx^\m\ot e_r \label{r47'}
\eeq
of $\gF$ onto $X$ is the strength form of a gauge potential $A$.

Let $I_k(e)=b_{r_1\ldots r_k}e^{r_1}\cdots e^{r_k}$ be a
$G$-invariant polynomial of degree $k>1$ on the Lie algebra $\cG$.
With $\gF$ (\ref{f34}), one can associate to $I_k$ the closed
gauge-invariant $2k$-form
\be
P_{2k}(\gF)=b_{r_1\ldots r_k}\gF^{r_1}\w\cdots\w \gF^{r_k}
\ee
on $C$. Given a section $B$ of $C\to X$, the pull-back $
P_{2k}(F_B)=B^*P_{2k}(\gF)$ of $P_{2k}(\gF)$ is a closed
characteristic form on $X$. Let the same symbol stand for its
pull-back onto $C$. Since $C\to X$ is an affine bundle and the de
Rham cohomology of $C$ equals that of $X$, the forms $P_{2k}(\gF)$
and $P_{2k}(F_B)$ possess the same cohomology class
$[P_{2k}(\gF)]=[P_{2k}(F_B)]$ for any principal connection $B$.
Thus, $I_k(e)\mapsto [P_{2k}(F_B)]\in H^*(X)$ is the familiar Weil
homomorphism. Furthermore, we obtain the transgression formula
\mar{r65}\beq
P_{2k}(\gF)-P_{2k}(F_B)=d\gS_{2k-1}(B), \qquad
\gS_{2k-1}(B)=k\op\int^1_0 \gP_{2k}(t,B)dt, \label{r65}
\eeq
on $C$ \cite{mpl}. Its pull-back by means of a section $A$ of
$C\to X$ gives the transgression formula
\be
P_{2k}(F_A)-P_{2k}(F_B)=d S_{2k-1}(A,B)
\ee
on $X$. For instance, if $P_{2k}(F_A)$ is the characteristic Chern
$2k$-form, then $S_{2k-1}(A,B)$ is the familiar Chern--Simons
$(2k-1)$-form. Therefore, we agree to call $\gS_{2k-1}(B)$
(\ref{r65}) the Chern--Simons form on the bundle $C$. Let us
consider the pull-back of this form onto the jet manifold $J^1C$
denoted by the same symbol $\gS_{2k-1}(B)$. Then $L_{CS}=h_0
\gS_{2k-1}(B)$ is the global Lagrangian of topological
Chern--Simons theory. One can show that its Lie derivative with
respect to any vector field $\xi$ (\ref{f33}) is $d_H$-exact
\cite{mpl}.

\section{Topological BF theory}

Let us consider topological BF theory of two exterior forms $A$
and $B$ of form degree $|A|+|B|=n-1$ on a smooth manifold $X$
\cite{birm}. It is a reducible $(n-3)$-stage degenerate Lagrangian
theory \cite{jpa05}. Since the verification of the homology
regularity condition in a general case is rather complicated, we
here restrict our consideration to the simplest example of the
topological BF theory when $A$ is a function
\cite{jmp05a,cmp06,jdg}.

Let us consider the fiber bundle $Y=\Bbb R\op\times_X \op\w^{n-1}
T^*X$, coordinated by $(x^\la, A, B_{\m_1\ldots \m_{n-1}})$ and
provided with the canonical $(n-1)$-form
\be
B=\frac{1}{(n-1)!}B_{\m_1\ldots \m_{n-1}}dx^{\m_1}\w\cdots\w
dx^{\m_{n-1}}.
\ee
The Lagrangian of topological BF theory reads
\mar{v182}\beq
L_{\rm BF}=\frac1n Ad_HB. \label{v182}
\eeq
Let us extend the original GDA $\cO^*_\infty Y$ of BF theory to
the GDA $\cP^*_\infty[\ol Y^*,Y]$ possessing the local basis $\{
A, B_{\m_1\ldots \m_{n-1}}, \ol s, \ol s^{\m_1\ldots \m_{n-1}}\}$,
where $\ol s, \ol s^{\m_1\ldots \m_{n-1}}$ are odd antifields of
antifield number 1. It is provided with the nilpotent Koszul--Tate
differential
\be
\ol\dl=\frac{\rdr}{\dr \ol s}\cE + \frac{\rdr}{\dr \ol
s^{\m_1\ldots \m_{n-1}}} \cE^{\m_1\ldots \m_{n-1}}.
\ee
Then one can show that the Noether identities (\ref{v64}) read
\mar{v191}\beq
\ol\dl \Delta^{\mu_2\ldots\mu_{n-1}}=
d_{\mu_1}\cE^{\mu_1\nu_2\ldots\mu_{n-1}}=0, \qquad
\Delta^{\mu_2\ldots\mu_{n-1}}= d_{\mu_1}\ol
s^{\mu_1\mu_2\ldots\mu_{n-1}}. \label{v191}
\eeq

The graded densities $\Delta^{\nu_2\ldots\nu_{n-1}}$ (\ref{v191})
form a local basis for a projective $C^\infty(X)$-module of finite
rank which, by virtue of the Serre--Swan theorem, is isomorphic to
the module of sections of the vector bundle
\be
\ol V^*=\op\w^{n-2} TX\op\ot_X \op\w^n T^*X, \qquad V= \op\w^{n-2}
T^*X.
\ee
Therefore, let us extend the GDA $\cP^*_\infty[\ol Y^*,Y]$ to the
BGDA $\cP^*_\infty\{0\}= \cP^*_\infty[\ol Y^*,Y,V]$ possessing the
local basis $\{A, B_{\m_1\ldots \m_{n-1}}, \ol s, \ol
s^{\m_1\ldots \m_{n-1}}, \ol c^{\m_2\ldots \m_{n-1}}\}$, where
$\ol c^{\m_2\ldots \m_{n-1}}$ are even antifields of antifield
number 2. We have the nilpotent graded derivation
\be
\dl_0= \ol\dl + \frac{\rdr}{\dr \ol c^{\m_2\ldots \m_{n-1}}}
\Delta^{\m_2\ldots \m_{n-1}}
\ee
of $\cP^*_\infty\{0\}$. Its nilpotency is equivalent to the
Noether identities (\ref{v191}).

Iterating the arguments, we come to the GDA $P^*\{n-2\}$
possessing the local basis
\be
\{A, B_{\m_1\ldots \m_{n-1}}, c_{\m_2\ldots \m_{n-1}},\ldots,
c_{\m_{n-1}},  c,
 \ol s, \ol s^{\m_1\ldots \m_{n-1}},
\ol c^{\m_2\ldots \m_{n-1}},\ldots,\ol c^{\m_{n-1}}, \ol c\},
\ee
where $\ol c^{\m_{k+2}\ldots \m_{n-1}}$, $k=0,\ldots,n-3$, are
antifields of Grassmann parity $(k+1){\rm mod}\,2$ and antifield
number $k+3$, $\ol c$ is an antifield of Grassmann parity
$(n-1){\rm mod}\,2$ and antifield number $n+1$, and $c_{\m_2\ldots
\m_{n-1}},\ldots, c_{\m_{n-1}},  c$ are the corresponding ghosts.
The GDA $P^*\{n-2\}$ is provided with the Koszul--Tate
differential
\be
&& \dl_{n-2}=\dl_0 + \op\sum_{1\leq k\leq n-3}\frac{\rdr}{\dr \ol
c^{\m_{k+2}\ldots \m_{n-1}}} \Delta^{\m_{k+2}\ldots \m_{n-1}} +
\frac{\rdr}{\dr \ol c}\Delta,\\
&& \Delta^{\m_{k+2}\ldots \m_{n-1}}=d_{\m_{k+1}}\ol
c^{\m_{k+1}\m_{k+2}\ldots \m_{n-1}}, \qquad \Delta=d_{\m_{n-1}}\ol
c^{\m_{n-1}}.
\ee
Its nilpotency results in the Noether identities (\ref{v191}) and
the $k$-stage Noether identities
\be
d_{\m_{k+2}}\Delta^{\m_{k+2}\ldots \m_{n-1}}=0, \qquad
k=0,\ldots,n-3.
\ee
By virtue of Noether's inverse second theorem, the total gauge
operator reads
\be
u_e=-d_{\m_1}c_{\m_2\ldots\m_{n-1}}\frac{\dr}{\dr
B_{\m_1\ldots\m_{n-1}}}- \op\sum_{1\leq k\leq n-3}
d_{\m_{k+1}}c_{\m_{k+2}\ldots\m_{n-1}}\frac{\dr}{\dr
c_{\m_{k+1}\ldots\m_{n-1}}} -d_\m c \frac{\dr}{\dr c_\m}
\ee
It is nilpotent and, thus, is the BRST operator. Accordingly, the
Lagrangian $L_{\rm BF}$ (\ref{v182}) is extended to a solution of
the master equation
\be
L_E= L_{\rm BF} + [c_{\m_2\ldots \m_{n-1}} d_{\m_1} \ol
s^{\m_1\m_2\ldots \m_{n-1}} + \op\sum_{1\leq k\leq n-3}
c_{\m_{k+2}\ldots \m_{n-1}} d_{\m_{k+1}} \ol
c^{\m_{k+1}\m_{k+2}\ldots \m_{n-1}} +c d_{\m_{n-1}}\ol
c^{\m_{n-1}}]\om.
\ee

\section{SUSY gauge theory}

SUSY gauge theory is mainly developed as Yang--Mills type theory
\cite{nill,wess,west}. However, its geometric formulation meets
difficulty because formalism of principal bundles in the
categories of graded manifolds and supermanifolds is rather
sophisticated \cite{bart,book00,shiggs,stavr}. ACFT overcomes this
difficulty \cite{ward,ward2}.

Let $\cG=\cG_0\oplus \cG_1$ be a finite-dimensional real Lie
superalgebra with a basis $\{e_r\}$, $r=1,\ldots,m,$ and real
structure constants $c^r_{ij}$. Recall that
\be
&& c^r_{ij}=-(-1)^{[i][j]}c^r_{ji}, \qquad [r]=[i]+[j],\\
&& (-1)^{[i][b]}c^r_{ij}c^j_{ab} + (-1)^{[a][i]}c^r_{aj}c^j_{bi} +
(-1)^{[b][a]}c^r_{bj}c^j_{ia}=0,
\ee
where $[r]$ denotes the Grassmann parity of $e_r$. Let the
modified structure constants
\mar{z90}\beq
\ol c^r_{ij}=(-1)^{[i]}c^r_{ij}, \qquad \ol
c^r_{ij}=(-1)^{([i]+1)([j]+1)}\ol c^r_{ji} \label{z90}
\eeq
be introduced. Given the universal enveloping algebra $\ol \cG$ of
$\cG$, we assume that there exists an invariant non-degenerate
even quadratic element $h^{ij}e_ie_j$ of $\ol\cG$.

In SUSY gauge theory on an Euclidean space $X=\Bbb R^n$, we
associate to the Lie superalgebra $\cG$ the GDA $\cP^*[Q;Y]$ where
\be
Q=X\times \cG_1)\op\ot_X T^*X, \qquad Y= (X\times \cG_0)\op\ot_X
T^*X.
\ee
Its basis is $(a^r_\la)$, $[a^r_\la]=[r]$. There is the canonical
decomposition
\be
a^r_{\la\m}=\frac12(\cF^r_{\la\m} +
\cS^r_{\la\m})=\frac12(a^r_{\la\m}-a^r_{\m\la} +c^r_{ij}a^i_\la
a^j_\m) +\frac12(a^r_{\la\m}+ a^r_{\m\la} -c^r_{ij}a^i_\la
a^j_\m).
\ee
Then the Yang--Mills graded Lagrangian takes the form
\be
L_{YM}=\frac14
h_{ij}\eta^{\la\m}\eta^{\bt\nu}\cF^i_{\la\bt}\cF^j_{\m\nu}\om,
\ee
where $\eta$ is an Euclidean metric on $\Bbb R^n$. Its variational
derivatives $\cE_r^\la$ obey the irreducible Noether identities
\be
 c^r_{ji}a^i_\la\cE_r^\la + d_\la\cE_j^\la=0.
\ee
Therefore, we enlarge the GDA $\cP^*[Q,Y]$ to the GDA $P^*\{0\}$
whose basis
\be
\{a^r_\la, c^r, \ol a^\la_r, \ol c_r\}, \qquad [c^r]=([r]+1){\rm
mod}\,2, \qquad [\ol a^\la_r]=[\ol c_r]=[r],
\ee
consists of gauge potentials $a^r_\la$, ghosts $c^r$ of ghost
number 1, and antifields $\ol a^\la_r$, $\ol c_r$ of antifield
numbers 1 and 2, respectively. Then the total gauge operator $u_e$
(\ref{w108}) reads
\be
u_e=u^r_\la\frac{\dr}{\dr a_\la^r}= (-c^r_{ji}c^ja^i_\la +
c^r_\la)\frac{\dr}{\dr a_\la^r}.
\ee
It admits the nilpotent BRST extension
\be
u_E=u_e +\xi= (-c^r_{ji}c^ja^i_\la + c^r_\la)\frac{\dr}{\dr
a_\la^r} -\frac12 \ol c^r_{ij}c^ic^j\frac{\dr}{\dr c^r},
\ee
where $\ol c^r_{ij}$ are the modified structure constants
(\ref{z90}). Then the Yang--Mills graded Lagrangian is extended to
a solution of the master equation
\be
L_E=L_{YM}+ (-c^r_{ij}c^ja^i_\la + c^r_\la)\ol a^\la_rd^nx
-\frac12 \ol c^r_{ij}c^ic^j\ol c_r\om.
\ee

\section{Field theory on composite bundles}

Let us consider a composite fiber bundle
\mar{1.34}\beq
Y\to \Si\to X, \label{1.34}
\eeq
where $\pi_{Y\Si}: Y\to\Si$ and $\pi_{\Si X}: \Si\to X$ are fiber
bundles. It is provided with fibered coordinates
$(x^\la,\si^m,y^i)$, where $(x^\m,\si^m)$ are bundle coordinates
on $\Si\to X$, i.e., the transition functions of coordinates
$\si^m$ are independent of coordinates $y^i$. The following facts
make composite bundles  useful for physical applications
\cite{book,book98,sard94,sau}.

Given a composite bundle (\ref{1.34}), let $h$ be a global section
of $\Si\to X$. Then the restriction
\mar{S10}\beq
Y_h=h^*Y \label{S10}
\eeq
of the fiber bundle $Y\to\Si$ to $h(X)\subset \Si$ is a subbundle
of the fiber bundle $Y\to X$.

Every section $s$ of the fiber bundle $Y\to X$ is a composition of
the section $h=\pi_{Y\Si}\circ s$ of the fiber bundle $\Si\to X$
and some section of the fiber bundle $Y\to \Si$ over $h(X)\subset
\Si$.

Let $J^1\Si$, $J^1_\Si Y$, and $J^1Y$ be jet manifolds of the
fiber bundles $\Si\to X$, $Y\to \Si$ and $Y\to X$, respectively.
They are provided with the adapted coordinates $(x^\la ,\si^m,
\si^m_\la)$, $( x^\la ,\si^m, y^i, \wt y^i_\la, y^i_m)$ and
$(x^\la ,\si^m, y^i, \si^m_\la ,y^i_\la)$. There is the canonical
map
\be
\vr : J^1\Si\op\times_\Si J^1_\Si Y\ar_Y J^1Y, \qquad
y^i_\la\circ\vr=y^i_m{\si}^m_{\la} +\wt y^i_{\la}.
\ee
Due to this map, any pair of connections
\mar{b1.113}\ben
&& A_\Si=dx^\la\ot (\dr_\la + A_\la^i\dr_i) +d\si^m\ot (\dr_m +
A_m^i\dr_i), \label{b1.113} \\
&& \G=dx^\la\ot (\dr_\la + \G_\la^m\dr_m) \nonumber
\een
on fiber bundles $Y\to \Si$ and $\Si\to X$, respectively,  yields
the composite connection
\mar{b1.114}\beq
\g=A_\Si\circ\G=dx^\la\ot (\dr_\la +\G_\la^m\dr_m + (A_\la^i +
A_m^i\G_\la^m)\dr_i) \label{b1.114}
\eeq
on the fiber bundle $Y\to X$. For instance, let us consider a
vector field $\tau$ on the base $X$, its horizontal lift $\G\tau$
onto $\Si$ by means of the connection $\G$ and, in turn, the
horizontal lift $A_\Si(\G\tau)$ of $\G\tau$ onto $Y$ by means of
the connection $A_\Si$. Then $A_\Si(\G\tau)$ is the horizontal
lift of $\tau$ onto $Y$ by means of the composite connection $\g$
(\ref{b1.114}).

Given a composite bundle $Y$ (\ref{1.34}), there is the exact
sequence of bundles
\mar{63}\beq
0\to V_\Si Y\to VY\to Y\op\times_\Si V\Si\to 0, \label{63}
\eeq
where $V_\Si Y$ is the vertical tangent bundle of the fiber bundle
$Y\to\Si$. Every connection $A$ (\ref{b1.113}) on the fiber bundle
$Y\to\Si$ yields the splitting
\be
\dot y^i\dr_i + \dot\si^m\dr_m= (\dot y^i -A^i_m\dot\si^m)\dr_i +
\dot\si^m(\dr_m+A^i_m\dr_i)
\ee
of the exact sequences (\ref{63}). This splitting defines the
first order differential operator
\mar{7.10}\beq
\wt D= dx^\la\otimes(y^i_\la- A^i_\la -A^i_m\si^m_\la)\dr_i
\label{7.10}
\eeq
on the composite bundle $Y\to X$. This operator, called the
vertical covariant differential, possesses the following important
property. Let $h$ be a section of the fiber bundle $\Si\to X$ and
$Y_h$ the subbundle (\ref{S10}) of the composite bundle $Y\to X$.
Then the restriction of the vertical covariant differential $\wt
D$ (\ref{7.10}) to $J^1Y_h\subset J^1Y$ coincides with the
familiar covariant differential relative to the pull-back
connection
\mar{mos83}\beq
A_h=h^*A_\Si=dx^\la\ot[\dr_\la+((A^i_m\circ h)\dr_\la h^m +(A\circ
h)^i_\la)\dr_i] \label{mos83}
\eeq
on $Y_h\to X$  \cite{book,book00,sard94}.

The peculiarity of field theory on a composite bundle (\ref{1.34})
is that its Lagrangian depends on a connection on $Y\to\Si$, but
not $Y\to X$, and it factorizes through the vertical covariant
differential (\ref{7.10}). This is the case of field theories with
broken symmetries, spinor fields, gauge gravitation theory
\cite{book00,sard98a,sard02,sard06,sard06a} and mechanical models
with parameters \cite{jmp02,jmp04,book05,book98,sard00}.

\section{Symmetry
 breaking and Higgs fields}

In gauge theory on a principal bundle $P\to X$, a symmetry
breaking is defined as reduction of the structure Lie group $G$ of
this principal bundle to a closed (consequently, Lie) subgroup $H$
of exact symmetries \cite{castr,book,keyl,nik,sard92,sard06a}.
From the mathematical viewpoint, one speaks on the Klein--Chern
geometry or a reduced $G$-structure \cite{gor,kob72,zul}.

By virtue of the well-known theorem \cite{kob,ste},  reduction of
the structure group of a principal bundle takes place iff there
exists a global section $h$ of the quotient bundle $P/H\to X$.
This section  is treated as a Higgs field. Thus, we have the
composite bundle
\mar{b3223a}\beq
P\to P/H\to X, \label{b3223a}
\eeq
where $P\to P/H$ is a principal bundle with the structure group
$H$ and  $\Si=P/H\to X$ is a $P$-associated fiber bundle with the
typical fiber $G/H$. Moreover, there is one-to-one correspondence
between the global sections $h$ of $\Si\to X$ and reduced
$H$-principal subbundles $P^h=\pi_{P\Si}^{-1}(h(X))$ of $P$.

Let $Y\to \Si$ be a vector bundle associated to the $H$-principal
bundle $P\to \Si$. Then sections of the composite bundle $Y\to
\Si\to X$ describe matter fields with the exact symmetry group $H$
in the presence of Higgs fields. Given bundle coordinates
$(x^\la,\si^m,y^i)$ on $Y$, these sections are locally represented
by  pairs $(\si^m(x), y^i(x))$. Given a global section $h$ of
$\Si\to X$, sections of the vector bundle $Y_h$ (\ref{S10})
describe matter fields in the presence of the background Higgs
field $h$. Moreover, for different Higgs fields $h$ and $h'$, the
fiber bundles $Y_h$ and $Y_{h'}$ need not be equivalent
\cite{book,sard92,sard06a}.

Note that $Y\to X$ fails to be associated to a principal bundle
$P\to X$ with the structure group $G$ and, consequently, it need
not admit a principal connection. Therefore, one should consider a
principal connection (\ref{b1.113}) on the fiber bundle $Y\to
\Si$, and a Lagrangian on $J^1Y$ factorizes through the vertical
covariant differential $\wt D$ (\ref{7.10}). In the presence of a
background Higgs field $h$, the restriction of $\wt D$ to $J^1Y_h$
coincides with the covariant differential relative to the
pull-back connection (\ref{mos83}) on $Y_h\to X$.

Riemannian and pseudo-Riemannian metrics on a manifold $X$
exemplify classical Higgs fields. Let $X$ be an oriented
four-dimensional smooth manifold and $LX$ the fiber bundle of
linear frames in the tangent spaces to $X$. It is a principal
bundle with the structure group $GL_4=GL^+(4,\Bbb R)$. By virtue
of the well known theorem \cite{ste}, this structure group is
always reducible to its maximal compact subgroup $SO(4)$. The
corresponding global sections of the quotient bundle $LX/SO(4)$
are Riemannian metrics on $X$. However, the reduction of the
structure group $GL_4$ of $LX$ to its Lorentz subgroup $SO(1,3)$
need not exist, unless $X$ satisfies certain topological
conditions. The quotient bundle
\mar{b3203}\beq
\Si_{\rm PR}= LX/SO(1,3)\to X, \label{b3203}
\eeq
is a natural bundle (see item 20), associated to $LX$. Its global
section $h$, called a tetrad field, defines a principal Lorentz
subbundle $L^hX$ of $LX$. Therefore, $h$ can be represented by a
family of local sections $\{h_a\}_\iota$ of $LX$ on trivialization
domains $U_\iota$ which take values in $L^hX$ and possess Lorentz
transition functions. One calls $\{h_a\}$ the tetrad functions or
vielbeins. They define an atlas $\Psi^h=\{(\{h_a\}_\iota,
U_\iota)\}$ of $LX$ and associated bundles with Lorentz transition
functions. There is the canonical imbedding of the bundle
$\Si_{\rm PR}$ (\ref{b3203}) onto an open subbundle of the tensor
bundle $\op\vee^2T^*X$ such that its global section $h=g$ is a
pseudo-Riemannian metric $g_{\m\nu}=h^a_\m h^b_\nu\eta_{ab}$ on
$X$. This fact motivates us to treat a metric (or tetrad)
gravitational field as a Higgs field
\cite{iva,sardz,sard02,sard06}.

Note that, if $G=GL_4$ and $H=SO(1,3)$, we are in the case of so
called reductive $G$-structure \cite{godina03} when the Lie
algebra $\ccG$ of $G$ is the direct sum
\mar{g13}\beq
{\cal G} = {\got h} \oplus {\got m} \label{g13}
\eeq
of the Lie algebra ${\got h}$ of $H$ and a subspace ${\got
m}\subset \ccG$ such that $ad(g)({\got m})\subset {\got m}$, $g\in
H$. In this case, the pull-back of the ${\got h}$-valued component
of any principal connection on $P$ onto a reduced subbundle $P^h$
is a principal connection on $P^h$.

\section{Dirac spinor fields}

Dirac spinors as like as other ones are described in the Clifford
algebra terms \cite{fried,law}. The Dirac spinor structure on a
four-dimensional manifold $X$ is defined as a pair $(P^h, z_s)$ of
a principal bundle $P^h\to X$ with the structure spin group
$L_s=SL(2,\Bbb C)$ and its bundle morphism $z_s: P^h \to LX$ to
the frame bundle $LX$ \cite{avis,law}. Any such morphism
factorizes
\mar{g10}\beq
P^h \to L^hX\to LX \label{g10}
\eeq
through some reduced principal subbundle $L^hX\subset LX$ with the
structure proper Lorentz group L$=SO^\uparrow(1,3)$, whose
universal two-fold covering is $\rL_s$. The corresponding quotient
bundle $\Si_{\rm T}=LX/\rL$ is a two-fold covering of the bundle
$\Si_{\rm PR}$ (\ref{b3203}). Its global sections are $L$-valued
tetrad fields $h$. Thus, any Dirac spinor structure is associated
to a Lorentz reduced structure, but the converse need not be true.
There is the well-known topological obstruction to the existence
of a Dirac spinor structure. For instance, a Dirac spinor
structure on a non-compact manifold $X$ exists iff $X$ is
parallelizable.

Given a Dirac spinor structure (\ref{g10}), the associated Dirac
spinor bundle $S^h$ can be seen as a subbundle of the bundle of
Clifford algebras generated by the Lorentz frames $\{t_a\}\in
L^hX$ \cite{benn,law}. This fact enables one to define the
Clifford representation
\mar{g11}\beq
\g_h(dx^\m)=h^\m_a\g^a \label{g11}
\eeq
of coframes $dx^\m$ in the cotangent bundle $T^*X$ by Dirac's
matrices, and introduce the Dirac operator on $S^h$ with respect
to a principal connection on $P^h$. Then sections of a spinor
bundle $S^h$ describe Dirac spinor fields in the presence of a
tetrad field $h$. However, the representations (\ref{g11}) for
different tetrad fields fail to be equivalent. Therefore, one
meets a problem of describing Dirac spinor fields in the presence
of different tetrad fields and under general covariant
transformations.

In order to solve this problem, let us consider the universal
two-fold covering $\wt{GL}_4$ of the group $GL_4$ and the
$\wt{GL}_4$-principal bundle $\wt{LX}\to X$ which is the two-fold
covering bundle  of the frame bundle $LX$ {\cite{dabr,law,swit}.
Then we have the commutative diagram
\be
\begin{array}{ccc}
 \wt{LX} & \ar^\zeta & LX \\
 \put(0,-10){\vector(0,1){20}} &
& \put(0,-10){\vector(0,1){20}}  \\
P^h & \ar & L^hX
\end{array}
\ee
for any Dirac spinor structure (\ref{g10})
\cite{fulp94,sard98a,sard02}. As a consequence,
$\wt{LX}/\rL_s=LX/L=\Si_{\rm T}$. Since $\wt{LX}\to \Si_T$ is an
$\rL_s$-principal bundle, one can consider the associated spinor
bundle $S\to \Si_T$ whose typical fiber is a Dirac spinor space
$V_s$ \cite{book00,sard98a,sard02}. We agree to call it the
universal spinor bundle because, given a tetrad field $h$, the
pull-back $S^h=h^*S\to X$ of $S$ onto $X$ is a spinor bundle on
$X$ which is associated to the $\rL_s$-principal bundle $P^h$. The
universal spinor bundle $S$ is endowed with bundle coordinates
$(x^\la, \si^\m_a,y^A)$, where $(x^\la, \si^\m_a)$ are bundle
coordinates on $\Si_T$ and $y^A$ are coordinates on the spinor
space $V_s$. The universal spinor bundle $S\to\Si_T$ is a
subbundle of the bundle of Clifford algebras which is generated by
the bundle of Minkowski spaces associated to the L-principal
bundle $LX\to\Si_T$. As a consequence, there is the Clifford
representation
\mar{L7}\beq
\g_\Si: T^*X\op\ot_{\Si_T} S \to S, \qquad \g_\Si (dx^\la)
=\si^\la_a\g^a, \label{L7}
\eeq
whose restriction to the subbundle $S^h\subset S$ restarts the
representation (\ref{g11}).

Sections of the composite bundle $S\to \Si_{\rm T}\to X$ describe
Dirac spinor fields in the presence of different tetrad fields as
follows \cite{sard98a,sard02}. Due to the splitting (\ref{g13}),
any general linear connection $K$ on $X$ (i.e., a principal
connection on $LX$) yields the connection
\mar{b3266}\ben
&& A_\Si = dx^\la\ot(\dr_\la - \frac14
(\eta^{kb}\si^a_\m-\eta^{ka}\si^b_\m)
 \si^\nu_k K_\la{}^\m{}_\nu L_{ab}{}^A{}_By^B\dr_A) +
 \label{b3266}\\
&& \qquad d\si^\m_k\ot(\dr^k_\m + \frac14 (\eta^{kb}\si^a_\m
-\eta^{ka}\si^b_\m) L_{ab}{}^A{}_By^B\dr_A) \nonumber
\een
on the universal spinor bundle $S\to\Si_T$. Its restriction to
$S^h$ is the familiar spin connection
\mar{b3212}\beq
K_h=dx^\la\ot[\dr_\la +\frac14
(\eta^{kb}h^a_\m-\eta^{ka}h^b_\m)(\dr_\la h^\m_k - h^\nu_k
K_\la{}^\m{}_\nu)L_{ab}{}^A{}_B y^B\dr_A], \label{b3212}
\eeq
defined by $K$ \cite{pon,sard97b}. The connection (\ref{b3266})
yields the vertical covariant differential
\mar{7.10'}\beq
\wt D= dx^\la\ot[y^A_\la- \frac14(\eta^{kb}\si^a_\m
-\eta^{ka}\si^b_\m)(\si^\m_{\la k} -\si^\nu_k
K_\la{}^\m{}_\nu)L_{ab}{}^A{}_By^B]\dr_A, \label{7.10'}
\eeq
on the fiber bundle $S\to X$. Its restriction to $J^1S^h\subset
J^1S$ recovers the familiar covariant differential on the spinor
bundle $S^h\to X$ relative to the spin connection (\ref{b3212}).
Combining (\ref{L7}) and (\ref{7.10'}) gives the first order
differential operator
\be
\cD=\si^\la_a\g^{aB}{}_A[y^A_\la- \frac14(\eta^{kb}\si^a_\m
-\eta^{ka}\si^b_\m)(\si^\m_{\la k} -\si^\nu_k
K_\la{}^\m{}_\nu)L_{ab}{}^A{}_By^B],
\ee
on the fiber bundle $S\to X$. Its restriction to $J^1S^h\subset
J^1S$ is the familiar Dirac operator on the spinor bundle $S^h$ in
the presence of a background tetrad field $h$ and a general linear
connection $K$.

\section{Natural and gauge-natural bundles}

A connection $\G$ on a fiber bundle $Y\to X$ defines the
horizontal lift $\G\tau$ onto $Y$ of any vector field $\tau$ on
$X$. There is the category of natural bundles \cite{kol,terng}
which admit the functorial lift $\wt\tau$ onto $T$ of any vector
field $\tau$ on $X$ such that $\tau\mapsto\ol\tau$ is a
monomorphism of the Lie algebra of vector field on $X$ to that on
$T$. One can think of the lift $\wt\tau$ as being an infinitesimal
generator of a local one-parameter group of general covariant
transformations of $T$. The corresponding Noether current
$\gJ_{\wt\tau}$ is the energy-momentum flow along $\tau$
\cite{giacqg,book,sard97b,sard98a}.

Natural bundles are exemplified by tensor bundles over $X$.
Moreover, all bundles associated to the principal frame bundle
$LX$ are natural bundles. The bundle
\mar{gr14}\beq
C_K=J^1LX/GL_4 \label{gr14}
\eeq
of principal connections on $LX$ is not associated to $LX$, but it
is also a natural bundle \cite{book,book00}. As is well known, a
spinor bundle $S^h$ associated to the spinor structure (\ref{g10})
fails to be a natural bundle. There exists the lift of any vector
field on $X$ onto $S^h$. It is called Kosmann's Lie derivative
\cite{fat98,god,kosm}. Such a lift is a property of any reductive
$G$-structure \cite{godina03}, but it is not a generator of
general covariant transformations. At the same time, the universal
spinor bundle $S\to X$ associated to the two-fold covering $\wt
LX$ of $LX$ is a natural bundle. Therefore, there exists the
functorial lift onto $S$ of any vector field on $X$. Its
restriction to a spinor bundle $S^h$ coincides with Kosmann's Lie
derivative \cite{sard97b,sard98a,sard02}.

In a more general setting, higher order natural bundles and
gauge-natural bundles are called into play
\cite{eck,fat03,kol,terng}. Note that the linear frame bundle $LX$
over a manifold $X$ is the set of first order jets of local
diffeomorphisms of $\Bbb R^n$ to $X$, $n=\di X$, at the origin of
$\Bbb R^n$. Accordingly, one considers $r$-order frame bundles
$L^rX$ of $r$-order jets of local diffeomorphisms of $\Bbb R^n$ to
$X$. Furthermore, given a principal bundle $P\to X$ with a
structure group $G$, the $r$-order jet bundle $J^1P\to X$ of its
sections fails to be a principal bundle. However, the product
$W^rP=L^rX\times J^rP$ is a principal bundle with the structure
group $W^r_nG$ which is a semi direct product of the group $G^r_n$
of invertible $r$-order jets of maps $\Bbb R^n$ to itself at its
origin (e.g., $G^1_n=GL(n,\Bbb R)$) and the group $T^r_nG$ of
$r$-order jets of morphisms $\Bbb R^n\to G$ at the origin of $\Bbb
R^n$. Moreover, if $Y\to X$ is a fiber bundle associated to $P$,
the jet bundle $J^rY\to X$ is a vector bundle associated to the
principal bundle $W^rP$. It exemplifies gauge natural bundles,
which can described as fiber bundles associated to principal
bundles $W^rP$. Natural bundles are gauge natural bundles for a
trivial $G=1$. The bundle of principal connections $C$ (\ref{f30})
is a first order gauge natural bundle. This fact motivates
somebody to develop generalized gauge theory on gauge natural
bundles \cite{eck,fat03,fran}.

\section{Gauge gravitation theory}

Gauge gravitation theory (see \cite{heh,iva,sardz,sard06} for a
survey) is described as a field theory on natural bundles over an
oriented four-dimensional manifold $X$ whose dynamic variables are
linear connections and pseudo-Riemannian metrics on $X$
\cite{ijgmmp05,fat04,book,sard97b,sard02,vign}.

Linear connections on $X$ (henceforth world connection) are
principal connections on the linear frame bundle $LX$  of $X$.
They are represented by sections of the bundle of linear
connections $C_K$ (\ref{gr14}). This is provided with bundle
coordinates $(x^\la,k_\la{}^\nu{}_\al)$ such that components
$k_\la{}^\nu{}_\al\circ K=K_\la{}^\nu{}_\al$ of a section $K$ of
$C_K\to X$ are coefficient of the linear connection
\be
K=dx^\la\ot (\dr_\la + K_\la{}^\m{}_\nu \dot x^\nu\dot\dr_\mu)
\ee
on $TX$ with respect to the holonomic bundle coordinates
$(x^\la,\dot x^\la)$. The first order jet manifold $J^1C_K$ of
$C_K$ admits the canonical decomposition taking the coordinate
form
\be
&& k_{\la\m}{}^\al{}_\bt=\frac12(R_{\la\m}{}^\al{}_\bt +
S_{\la\m}{}^\al{}_\bt)=\frac12(k_{\la\m}{}^\al{}_\bt -
k_{\m\la}{}^\al{}_\bt + k_\m{}^\al{}_\ve k_\la{}^\ve{}_\bt
-k_\la{}^\al{}_\ve k_\m{}^\ve{}_\bt)+  \\
&& \qquad \frac12(k_{\la\m}{}^\al{}_\bt + k_{\m\la}{}^\al{}_\bt -
k_\m{}^\al{}_\ve k_\la{}^\ve{}_\bt +k_\la{}^\al{}_\ve
k_\m{}^\ve{}_\bt).
\ee
If $K$ is a section of $C_K\to X$, then $R\circ K$ is the
curvature of a world connection $K$.

In order to describe gravity, let us assume that the linear frame
bundle $LX$ admits a Lorentz structure, i.e., reduced principal
subbundles with the structure Lorentz group. Sections of the
corresponding quotient bundle $\Si_{\rm PR}$ (\ref{b3203}) are
pseudo-Riemannian (henceforth world) metrics on $X$. Note that the
physical underlying reasons for the existence of a Lorentz
structure and, consequently, a world metric are both the geometric
equivalence principle and the existence of Dirac fermion fields
\cite{iva,sardz,sard02}.

The total configuration space of gauge gravitation theory in the
absence of matter fields is the bundle product $\Si_{PR}\times
C_K$ coordinated by $(v^\la,\si^{\al\bt},  k_\mu{}^\al{}_\bt)$.
This is a natural bundle admitting the functorial lift
\mar{gr3}\ben
&& \wt\tau_{K\Si}=\tau^\m\dr_\m +(\si^{\nu\bt}\dr_\nu \tau^\al
+\si^{\al\nu}\dr_\nu \tau^\bt)\frac{\dr}{\dr \si^{\al\bt}} +
\label{gr3}\\
&& \qquad (\dr_\nu \tau^\al k_\m{}^\nu{}_\bt -\dr_\bt \tau^\nu
k_\m{}^\al{}_\nu -\dr_\mu \tau^\nu k_\nu{}^\al{}_\bt
+\dr_{\m\bt}\tau^\al)\frac{\dr}{\dr k_\mu{}^\al{}_\bt} \nonumber
\een
of vector fields $\tau$ on $X$ \cite{ijgmmp05,book00}. These lifts
are generators of one-dimensional groups of general covariant
transformations, whose gauge parameters are vector fields on $X$.

We do not specify a gravitation Lagrangian $L_G$ on the jet
manifold $J^1(\Si_{PR}\times C_K)$, but assume that vector fields
(\ref{gr3}) exhaust its gauge symmetries. Then the Euler--Lagrange
operator $(\cE_{\al\bt} d\si^{\al\bt} + \cE^\m{}_\al{}^\bt
dk_\m{}^\al{}_\bt)\w\om$ of this Lagrangian obeys irreducible
Noether identities
\be
&&-(\si^{\al\bt}_\la +2\si^{\nu\bt}_\nu\dl^\al_\la)\cE_{\al\bt}
-2\si^{\nu\bt}d_\nu\cE_{\la\bt} +(-k_{\la\m}{}^\al{}_\bt
-k_{\nu\m}{}^\nu{}_\bt\dl^\al_\la + k_{\bt\m}{}^\al{}_\la +
k_{\m\la}{}^\al{}_\bt)\cE^\m{}_\al{}^\bt +\\
&& \qquad (-k_\m{}^\nu{}_\bt\dl^\al_\la
+k_\m{}^\al{}_\la\dl^\nu_\bt
+k_\la{}^\al{}_\bt\dl^\nu_\m)d_\nu\cE^\m{}_\al{}^\bt + d_{\m\bt}
\cE^\m{}_\la{}^\bt=0
\ee
\cite{ijgmmp05}. Taking the vertical part of vector fields
$\wt\tau_{K\Si}$ and replacing gauge parameters $\tau^\la$ with
ghosts $c^\la$, we obtain the total gauge operator and its
nilpotent BRST prolongation
\be
&&u_E=u^{\al\bt}\frac{\dr}{\dr\si^{\al\bt}} +u_\m{}^\al{}_\bt
\frac{\dr}{\dr k_\mu{}^\al{}_\bt} +u^\la \frac{\dr}{\dr
c^\la}=(\si^{\nu\bt} c_\nu^\al +\si^{\al\nu}
c_\nu^\bt-c^\la\si_\la^{\al\bt})\frac{\dr}{\dr \si^{\al\bt}}+
\\
&& \qquad (c_\nu^\al k_\m{}^\nu{}_\bt -c_\bt^\nu k_\m{}^\al{}_\nu
-c_\mu^\nu k_\nu{}^\al{}_\bt +c_{\m\bt}^\al-c^\la
k_{\la\mu}{}^\al{}_\bt)\frac{\dr}{\dr k_\mu{}^\al{}_\bt} +
c^\la_\m c^\m\frac{\dr}{\dr c^\la},
\ee
but this differs from that in \cite{gron}. Accordingly, an
original Lagrangian $L_G$ is extended to a solution of the master
equation
\be
L_E= L_G + u^{\al\bt}\ol\si_{\al\bt}\om + u_\m{}^\al{}_\bt \ol
k^\m{}_\al{}^\bt\om + u^\la \ol c_\la\om,
\ee
where $\ol\si_{\al\bt}$, $\ol k^\m{}_\al{}^\bt$ and $\ol c_\la$
are corresponding antifields.

\section{Covariant Hamiltonian field theory}

As is well-known, the familiar symplectic Hamiltonian technique
applied to field theory leads to instantaneous Hamiltonian
formalism on an infinite-dimensional phase space coordinated by
field functions at some instant of time \cite{got91a}. The true
Hamiltonian counterpart of classical first order Lagrangian field
theory on a fiber bundle $Y\to X$ is covariant Hamiltonian
formalism, where canonical momenta $p^\m_i$ correspond to
derivatives $y^i_\m$ of field variables $y^i$ with respect to all
world coordinates $x^\m$.  This formalism has been rigorously
developed since the 1970s in the multisymplectic, polysymplectic
and Hamilton -- De Donder variants (see
\cite{ech00,ech04,for1,book,jpa99,poly,got91,hel,krupk1,leon,leon02,leon05,marsd,palese,rey}
and references therein).

The multisymplectic phase space is the homogeneous Legendre bundle
\mar{N41}\beq
Z_Y= T^*Y\w(\op\w^{n-1}T^*X), \label{N41}
\eeq
coordinated by $(x^\la,y^i,p^\la_i,p)$. It is endowed with the
canonical exterior form
\be
\Xi_Y= p\om + p^\la_i dy^i\w\om_\la,
\ee
whose exterior differential $d\Xi_Y$ is the multisymplectic form
\cite{cantr,mart}. Given a first order Lagrangian $L=\cL\om$ on
$J^1Y$, the associated Poincar\'e--Cartan form
\mar{303}\beq
 H_L=L +\pi^\la_i\th^i\w\om_\la, \quad \pi^\la_i=\dr^\la_i\cL,
 \quad \om_\la=\dr_\la\rfloor\om.
\label{303}
\eeq
is a Lepagean equivalent both of $L$ and the Lagrangian
\mar{cmp80}\beq
 \ol L=\wh h_0(H_L) = (\cL + (\wh y_\la^i -
y_\la^i)\pi_i^\la)\om, \qquad \wh h_0(dy^i)=\wh y^i_\la dx^\la,
\label{cmp80}
\eeq
on the repeated jet manifold $J^1J^1Y$, whose Euler--Lagrange
equations are the Cartan ones
\mar{b336}\beq
\dr_i^\la\pi_j^\m(\wh y_\m^j - y_\m^j)=0, \quad
 \dr_i \cL - \wh
d_\la\pi_i^\la + (\wh y_\la^j - y_\la^j)\dr_i\pi_j^\la=0.
\label{b336}
\eeq
The Poincar\'e--Cartan form (\ref{303}) yields the Legendre
morphism
\be
\wh H_L: J^1Y\op\to_Y Z_Y, \quad (p^\m_i, p)\circ\wh H_L
=(\pi^\m_i, \cL-\pi^\m_i y^i_\m ),
\ee
of $J^1Y$ to the homogeneous Legendre bundle $Z_Y$. If its image
$Z_L=\wh H_L(J^1Y)$ is an imbedded subbundle $i_L:Z_L\to Z_Y$ of
$Z_Y\to Y$, it is provided with the pull-back De Donder form
$\Xi_L=i^*_L\Xi_Y$. The Hamilton -- De Donder equations for
sections $\ol r$ of $Z_L\to X$ read
\mar{N46}\beq
\ol r^*(u\rfloor d\Xi_L)=0, \label{N46}
\eeq
where $u$ is an arbitrary vertical vector field on $Z_L\to X$. If
the Legendre morphism $\wh H_L$ is a submersion, one can show that
a section $\ol s$ of $J^1Y\to X$ obeys the Cartan equations
(\ref{b336}) iff $\wh H_L\circ\ol s$ satisfies the Hamilton--De
Donder ones (\ref{N46}) \cite{jpa99,got91}. In a general setting,
one studies different Lepagean forms in order to develop Hamilton
-- De Donder formalism \cite{krupk1,krupk2}.

The homogeneous Legendre bundle $Z_Y$ is the trivial
one-dimensional bundle $\zeta:Z_Y\to \Pi$ over the Legendre bundle
\mar{00}\beq
\Pi=\op\w^nT^*X\op\ot_YV^*Y\op\ot_YTX=V^*Y\w(\op\w^{n-1}T^*X),
\label{00}
\eeq
coordinated by $(x^\la,y^i,p^\m_i)$. Being provided with the
canonical polysymplectic form
\be
\Om =dp_i^\la\w dy^i\w \om\ot\dr_\la,
\ee
the Legendre bundle $\Pi$ is the momentum phase space of
polysymplectic Hamiltonian formalism
\cite{cari91,book,jpa99,gunt,kij,sardz93,sard94a}.  A Hamiltonian
$\cH$ on $\Pi$ is defined as a section $p=-\cH$ of the bundle
$\zeta$. The pull-back of $\Xi_Y$ onto $\Pi$ by a $\cH$ is a
Hamiltonian form
\mar{b418}\beq
 H=\cH^*\Xi_Y= p^\la_i dy^i\w \om_\la -\cH\om. \label{b418}
\eeq
For every Hamiltonian form $H$ (\ref{b418}), there exists a
connection $\g$ on $\Pi\to X$ such that $\g\rfloor\Om= dH$. This
connection yields the first order Hamilton equations
\mar{b4100}\beq
y^i_\la=\dr^i_\la\cH, \qquad p^\la_{\la i}=-\dr_i\cH \label{b4100}
\eeq
on $\Pi$ which are exactly the Euler--Lagrange equations for the
first-order Lagrangian
\mar{m5}\beq
L_\cH=h_0(H)= (p^\la_i y^i_\la-\cH)\om \label{m5}
\eeq
on $J^1\Pi$. Let $i_N:N\to \Pi$ be a closed imbedded subbundle of
the Legendre bundle $\Pi\to Y$ which is regarded as a constraint
space. Let $H_N=i^*_NH$ be the pull-back of the Hamiltonian form
$H$ (\ref{b418}) onto $N$. This form defines the constrained
Lagrangian
\mar{cmp81}\beq
L_N=h_0(H_N)=(J^1i_N)^*L_\cH \label{cmp81}
\eeq
on the jet manifold $J^1N_L$. In fact, this Lagrangian is the
restriction of $L_\cH$ to $N\times J^1Y$. Its Euler--Lagrange
equations are called the constrained Hamilton equations. One can
show that any solution of the Hamilton equations (\ref{b4100})
which lives in the constraint manifold $N$ is also a solution of
the constrained Hamilton equations on $N$ \cite{book,jpa99}.

Lagrangian and covariant Hamiltonian formalisms are not
equivalent, unless Lagrangians are hyperregular. The key point is
that a non-regular Lagrangian admits different associated
Hamiltonians, if any. At the same time, there is a comprehensive
relation between these formalisms in the case of almost-regular
Lagrangians \cite{book,jpa99,sardz93}.

Any first order Lagrangian $L$ yields the Legendre map
\be
\wh L: J^1Y\ar_Y \Pi, \quad p^\la_i\circ\wh L=\dr^\la_i\cL,
\ee
whose image $N_L=\wh L(J^1Y)$ is called the Lagrangian constraint
space. Conversely, any Hamiltonian $\cH$ defines the Hamiltonian
map
\be
\wh H: \Pi\ar_Y J^1Y, \quad y_\la^i\circ\wh H=\dr^i_\la\cH.
\ee
A Hamiltonian $\cH$ on $\Pi$ is said to be associated to a
Lagrangian $L$ if it satisfies the relations
\be
p^\m_i=\dr^\m_i\cL (x^\m,y^i,\dr^j_\la\cH), \qquad
p^\m_i\dr^i_\m\cH-\cH=\cL(x^\m,y^j,\dr^j_\la\cH).
\ee
A Lagrangian $L$ is called almost-regular if the Lagrangian
constraint space $N_L$ is a closed imbedded subbundle of the
Legendre bundle $\Pi\to Y$, and $\wh L:J^1Y\to N_L$ is a fibered
manifold with connected fibers. In this case, the
Poincar\'e--Cartan form (\ref{303}) is the pull-back $H_L=\wh
L^*H$ of the Hamiltonian form $H$ (\ref{b418}) for any associated
Hamiltonian $\cH$. If an almost-regular Lagrangian admits
associated Hamiltonians $\cH$, they define a unique constrained
Lagrangian $L_N=h_0(H_N)$ (\ref{cmp81}) on the jet manifold
$J^1N_L$ of the fiber bundle $N_L\to X$. Then one can show that a
section $\ol s$ of the jet bundle $J^1Y\to X$ is a solution of the
Cartan equations for $L$ iff $\wh L\circ \ol s$ is a solution of
the constrained Hamilton equations.

For instance, the comprehensive description of systems with
almost-regular quadratic Lagrangians can be obtained
\cite{book,jpa99,poly}. In this case, the jet bundle $J^1Y\to Y$
admits a splitting similar to that (\ref{f31}) in gauge theory. As
a consequence, such a Lagrangian is brought into the Yang--Mills
type form, and can be accordingly quantized
\cite{bashk,bashk1,poly}.

In order to quantize covariant Hamiltonian field theory, one often
try to construct the multisymplectic generalization of a Poisson
bracket \cite{castr2,for,for2,hip,kanat,kanat2,kastrup,bracket}.
In a different way, we quantize covariant (polysymplectic)
Hamiltonian field theory as a particular Lagrangian system with
the Lagrangian $L_\cH$ (\ref{m5}) in path integral terms
\cite{bashk,bashk1,sard94b}.

There are attempts to generalize covariant Hamiltonian formalism
(e.g., its Hamilton--De Donder variant) to higher order Lagrangian
systems \cite{ald2,kru02,krupk2,snad}. However, a problem is to
define the Legendre map $\wh L$ for a higher order Lagrangian $L$.

\section{Time-dependent mechanics}

Non-relativistic time-dependent mechanics (see item 25 for
relativistic one) can be formulated as particular field theory on
fiber bundles $Q\to\Bbb R$ over a time axis $\Bbb R$
\cite{giach92,krupva,leon97a,book98,massa,sard98,sard00}. In this
case, polysymplectic and multisymplectic Hamiltonian formalisms
provide Hamiltonian and homogeneous Hamiltonian formulations of
time-dependent mechanics, whose momentum and homogeneous momentum
phase spaces are the vertical cotangent bundle $V^*Q$ of $Q\to\Bbb
R$ and the cotangent bundle $T^*Q$, respectively
\cite{book98,sard98,sard00}.

At the same time, there is the essential difference between field
theory and time-dependent mechanics. In contrast with gauge
potentials in field theory, connections on a configuration bundle
$Q\to\Bbb R$ of time-dependent mechanics fail to be dynamic
variables since their curvature vanishes. There is one-to-one
correspondence between these connections and the trivializations
$Q\ap \Bbb R\times M$ of a configuration space, i.e., reference
frames \cite{book98,book00,sard98}. If a reference frame holds
fixed, time-dependent mechanical systems are familiarly described
on the products $Q\ap \Bbb R\times M$, $J^1Q\ap \Bbb R\times TM$,
$V^*Q\ap \Bbb R\times T^*M$, which are not subject to
time-dependent transformations
\cite{cari89,cari93,chinea,eche,ham,mora}.

\section{Jets of submanifolds}

Jets of sections of fiber bundles are particular jets of
submanifolds. Namely, a space of jets of submanifolds admits a
cover by charts of jets of sections \cite{book,subm,kras,modu}.
Three-velocities in relativistic mechanics exemplify first order
jets of submanifolds (see item 25). A problem is that differential
forms on jets of submanifolds do not constitute a variational
bicomplex because horizontal forms (e.g., Lagrangians) are not
preserved under coordinate transformations. However, one can
associate to jets of $n$-dimensional submanifolds of an
$m$-dimensional manifold $Z$ the jets of sections of a trivial
fiber bundle
\mar{s12}\beq
Z_Q=Q\times Z\to Q, \label{s12}
\eeq
where $Q$ is some $n$-dimensional manifold. This relation fails to
be one-to-one correspondence. The ambiguity contains, e.g.,
diffeomorphisms of $Q$. Lagrangian formalism on a fiber bundle
(\ref{s12}) is developed in a standard way, but Lagrangians are
required to be variationally invariant under the above mentioned
diffeomorphisms of $Q$ (see item 26) \cite{subm}.

Given an $m$-dimensional smooth real manifold $Z$, a $k$-order jet
of $n$-dimensional submanifolds of $Z$ at a point $z\in Z$ is
defined as the equivalence class $j^k_zS$ of $n$-dimensional
imbedded submanifolds of $Z$ through $z$ which are tangent to each
other at $z$ with order $k$. The set $J^k_nZ$ of this jets is a
finite-dimensional real smooth manifold. Let $Y\to X$ be an
$m$-dimensional fiber bundle over an $n$-dimensional base $X$ and
$J^kY$ the $k$-order jet manifold of sections of $Y\to X$. Given
an imbedding $Y\to Z$, there is the natural injection $J^kY\to
J^k_nZ$ which defines a chart on $J^k_nZ$. These charts provide a
manifold atlas of $J^k_nZ$.

In particular, there is obvious one-to-one correspondence between
the jets $j^1_zS$ at a point $z\in Z$ and the $n$-dimensional
vector subspaces of the tangent space $T_zZ$ of $Z$ at $z$. It
follows that $J^1_nZ$ is a fiber bundle $\rho:J^1_nZ\to Z$ in
Grassmann manifolds. It possesses the following coordinate atlas.
Let $\{(U;z^\m)\}$ be a coordinate atlas of $Z$. Putting
$J^0_nZ=Z$, let us provide $J^0_mZ$ with the atlas obtained by
replacing every chart $(U,z^A)$ of $Z$ with the $m!/(n!(m-n)!)$
charts on $U$ which correspond to different partitions of $(z^A)$
in collections of $n$ and $m-n$ coordinates $(x^a,y^i)$,
$a=1,\ldots,n$, $i=1,\ldots,m-n$. Accordingly, the first order jet
manifold $J^1_nZ$ is endowed with the coordinates
$(x^a,y^i,y^i_a)$ possessing transition functions
\mar{s2}\beq
x'^a = x'^a (x^b, y^k), \quad y'^i = y'^i (x^b, y^k), \qquad
y'^j_a= ( \frac{\dr y'^j}{\dr y^k} y^k_b+ \frac{\dr y'^j}{\dr
x^b})(\frac{\dr x^b}{\dr y'^i} y'^i_a + \frac{\dr x^b}{\dr x'^a}).
\label{s2}
\eeq
In particular, if coordinate transition functions $x'^a$ are
independent of coordinates $y^k$, the transformation law
(\ref{s2}) comes to the familiar transformations of jets of
sections.

Given a coordinate chart $(\rho^{-1}(U);x^a,y^i,y^i_a)$ of
$J^1_nZ$, one can regard $\rho^{-1}(U)$ as the first order jet
manifold $J^1U$ of sections of the fiber bundle $U\ni (x^a,y^i)\to
(x^a)\in U_X$. The graded differential algebra
$\cO^*(\rho^{-1}(U))$ of exterior forms on $\rho^{-1}(U)$ is
generated by horizontal forms $dx^a$ and contact forms
$dy^i-y^i_adx^a$. Coordinate transformations (\ref{s2}) preserve
the ideal of contact forms, but not horizontal forms. Therefore,
one can develop first order Lagrangian formalism with a Lagrangian
$L=\cL d^nx$ on a coordinate chart $\rho^{-1}(U)$, but this
Lagrangian fails to be globally defined on $J^1_nZ$.

In order to overcome this difficulty, let us consider the above
mentioned bundle $Z_Q$ (\ref{s12}), coordinated by
$(q^\m,x^a,y^i)$, and its first order jet manifold $J^1Z_Q$
endowed with coordinates $(q^\m,x^a,y^i,x^a_\m, y^i_\m)$
possessing transition functions
\mar{s16}\ben
&& q'^\m=q^\m(q^\nu), \qquad x'^a = x'^a (x^b, y^k), \qquad y'^i =
y'^i (x^b, y^k), \label{s16}\\
&& x'^a_\m=(\frac{\dr x'^a}{\dr y^k}y^k_\nu + \frac{\dr x'^a}{\dr
x^b}x^b_\nu )\frac{\dr q^\nu}{\dr q'^\m}, \qquad
y'^i_\m=(\frac{\dr y'^i}{\dr y^k}y^k_\nu + \frac{\dr y'^i}{\dr
x^b}x^b_\nu)\frac{\dr q^\nu}{\dr q'^\m}. \nonumber
\een
An element $(q^\m,x^a,y^i,x^a_\m, y^i_\m)\in J^1Z_Q$ is called
regular if an $m\times n$ matrix with the entries $(x^a_\m,
y^i_\m)$ is of maximal rank $n$. This property is preserved under
the coordinate transformations (\ref{s16}). Obviously, any regular
elements of $J^1Z_Q$ defines some jet of $n$-dimensional
subbundles of the manifold $\{q\}\times Z$ through a point
$(x^a,y^i)\in Z$. Moreover, one can state the following relations
between the elements of $J^1_nZ$ and the regular elements of
$J^1Z_Q$ \cite{subm}.

Any jet of submanifolds $(x^a,y^i, y^i_a)$ through a point $z\in
Z$ defines some (but not unique) jet $(q^\m,x^a,y^i,x^a_\m,
y^i_\m)$ of sections of the fiber bundle $Z_Q$ (\ref{s12}) through
a point $q\times z$ for any $q\in Q$ if the jet coordinates obey
the equalities
\mar{s17}\beq
y^i_a x^a_\m = y^i_\m. \label{s17}
\eeq
Any regular element $(q^\m,x^a,y^i,x^a_\m, y^i_\m)$ of $J^1Z_Q$
defines a unique element $(x^a,y^i, y^i_a)$ of the jet manifold
$J^1_nZ$ by means of the equalities
\mar{s31}\beq
y^i_a=y^i_\m(x^{-1})^\m_a. \label{s31}
\eeq
The equalities (\ref{s17}) and (\ref{s31}) are maintained under
coordinate transformations (\ref{s2}) -- (\ref{s16}). Note that
there is a certain ambiguity between elements of $J^1_nZ$ and
$J^1Z_Q$. Non-regular elements of $J^1Z_Q$ can correspond to
different jets of submanifolds. Two regular elements
$(q^\m,z^A,z^A_\m)$ and $(q^\m,z^A,z'^A_\m)$ of $J^1Z_Q$ define
the same jet of submanifolds if $z'^A_\m= M^\nu_\mu z^A_\nu$,
where $M$ is some matrix. For instance, $M$ comes from a
diffeomorphism of $Q$.

Basing on this result, one can describe the dynamics of
$n$-dimensional submanifolds of a manifold $Z$ as that of sections
of the fiber bundle (\ref{s12}) (see item 26).

\section{Relativistic mechanics}

Given an $m$-dimensional manifold $Z$ coordinated by $(z^A)$, let
us consider the jet manifold $J^1_1Z$ of its one-dimensional
submanifolds. Let us provide $Z=J^0_1Z$ with coordinates
$(x^0=z^0, y^i=z^i)$. Then the jet manifold $J^1_1Z$ is endowed
with coordinates $(z^0,z^i,z^i_0)$ possessing transition functions
(\ref{s2}) which read
\mar{s120}\beq
z'^0=z'^0(z^0,z^k), \quad z'^0=z'^0(z^0,z^k), \quad z'^i_0=
(\frac{\dr z'^i}{\dr z^j} z^j_0 + \frac{\dr z'^i}{\dr z^0} )
(\frac{\dr z'^0}{\dr z^j} z^j_0 + \frac{\dr z'^0}{\dr z^0} )^{-1}.
\label{s120}
\eeq
A glance at this expression shows that $J^1_1Z\to Z$ is a fiber
bundle in projective spaces. For instance, put $Z=\Bbb R^4$ whose
Cartesian coordinates are subject to Lorentz transformations
\mar{s122}\beq
z'^0= z^0{\rm ch}\al - z^1{\rm sh}\al, \qquad z'^'= -z^0{\rm
sh}\al + z^1{\rm ch}\al, \qquad z'^{2,3} = z^{2,3}. \label{s122}
\eeq
Then $z'^i$ (\ref{s120}) are exactly the Lorentz transformations
\be
z'^1_0=\frac{ z^1_0{\rm ch}\al -{\rm sh}\al}{ - z^1_0{\rm sh}\al+
{\rm ch}\al} \qquad z'^{2,3}_0=\frac{z^{2,3}_0}{ - z^1_0{\rm
sh}\al + {\rm ch}\al}
\ee
of three-velocities in relativistic mechanics
\cite{subm,book98,sard98,sard03}.

Let us consider a one-dimensional manifold $Q=\Bbb R$ and the
product $Z_Q=\Bbb R\times Z$. Let $\Bbb R$ be provided with a
Cartesian coordinate $\tau$ possessing transition function
$\tau'=\tau + {\rm const}$. Then the jet manifold $J^1Z_Q$ of the
fiber bundle $\Bbb R\times Z\to \Bbb R$ is endowed with the
coordinates $(\tau, z^0, z^i, z^0_\tau, z^i_\tau)$ with the
transition functions
\be z'^0_\tau=\frac{\dr z'^0}{\dr z^k} z^k_\tau +
\frac{\dr z'^0}{\dr z^0} z^0_\tau, \qquad z'^i_\tau=\frac{\dr
z'^i}{\dr z^k} z^k_\tau + \frac{\dr z'^i}{\dr z^0} z^0_\tau.
\ee
In the case of Lorentz transformations (\ref{s122}), these
transition functions are transformations of four-velocities in
relativistic mechanics where $\tau$ is a proper time.

Let us consider coordinate charts $(U';\tau, z^0,z^i,z^i_0)$ and
$(U'';\tau, z^0,z^i,z^0_\tau, z^i_\tau)$ of the manifolds $\Bbb
R\times J^1_1Z$ and $J^1Z_Q$ over the same chart $(U;\tau,
z^0,z^i)$ of $Z_Q$. Then one can associate to each element $(\tau,
z^0,z^i,z^i_0)$ of $U'$ the elements of $U''$ which obey the
relations (\ref{s17})-- (\ref{s31}):
\mar{s125}\beq
z^i_0 z^0_\tau= z^i_\tau, \qquad  z^i_0=
\frac{z^i_\tau}{z^0_\tau}, \qquad z^0_\tau\neq 0. \label{s125}
\eeq
Given a point $(\tau,z)\in \Bbb R\times Z$, the relations
(\ref{s125}) are exactly the correspondence between elements of a
one-dimensional vector subspace of the tangent space $T_zZ$ and
the corresponding element of the projective space of these
subspaces. In relativistic mechanics, the relations (\ref{s125})
are familiar equalities between three- and four-velocities, and
one avoids the ambiguity between them by means of the nonholonomic
constraint $(z^0_\tau)^2- \op\sum_i(z^i_\tau)^2 =1$.

\section{String theory}

Given a manifold $Z$, one can develop Lagrangian theory of its
$n$-dimensional submanifolds as Lagrangian theory on the fiber
bundle $Z_Q$ (\ref{s12}) for an appropriate $n$-dimensional
manifold $Q$. If $n=2$, we are in the case of classical string
theory.

Let $Z_Q$ be a fiber bundle (\ref{s12}) coordinated by $(q^\m,
z^A)$ and $J^1Z_Q$ its first order jet manifold provided with
coordinates $(q^\m, z^A, z^A_\m)$, possessing transition functions
\be
q'^\m(q^\nu), \quad, z'^A(z^B), \qquad z'^A_\m=\frac{\dr z'^A}{\dr
z^B} \frac{\dr q^\nu}{\dr q'^\m} z^B_\nu.
\ee
Let $L=\cL(z^A, z^A_\m) d^nq$ be a first order Lagrangian on
$J^1Z_Q$ and $\dl L= \cE_A dz^A\w d^nq$ its Euler--Lagrange
operator. Let us consider an arbitrary vector field
$u=u^\m(q^\nu)\dr_\m$ on $Q$. It is an infinitesimal generator of
a one-parameter group of local diffeomorphisms of $Q$. Since
$Z_Q\to Q$ is a trivial bundle, this vector field gives rise to a
vector field $u=u^\m\dr_\m$ on $Z_Q$, and its jet prolongation
onto $J^1Z_Q$ reads
\mar{s59}\beq
u= u^\m \dr_\m - z^A_\nu\dr_\m u^\nu \dr_A^\m= u^\m d_\m +[- u^\nu
z^A_\nu\dr_A - d_\m (u^\nu z^A_\nu)\dr_A^\m]. \label{s59}
\eeq
One can regard it as a generalized vector field depending on
parameter functions $u^\m(q^\nu)$. In order to describe jets of
submanifolds of $Z$, it seems reasonable to require that a
Lagrangian $L$ on $J^1Z_Q$ is independent on coordinates of $Q$
and variationally invariant under $u$ (\ref{s59}) or,
equivalently, its vertical part
\be
u_V= - u^\nu z^A_\nu\dr_A - d_\m(u^\nu z^A_\nu)\dr_A^\m.
\ee
Then the variational derivatives of this Lagrangian obey
irreducible Noether identities
\mar{s60}\beq
z^A_\nu\cE_A=0. \label{s60}
\eeq

For instance, let us consider Lagrangian theory of two-dimensional
submanifolds (strings) \cite{subm}. Let $Z$ be an $m$-dimensional
locally affine manifold, i.e., a toroidal cylinder $\Bbb
R^{m-k}\times T^k$. Its tangent bundle $TZ$ can be provided with a
constant non-degenerate fiber metric $\eta_{AB}$. Let $Q$ be a
two-dimensional manifold. Let us consider the $2\times 2$ matrix
with the entries $h_{\m\nu}=\eta_{AB} z^A_\m z^B_\nu$. Then its
determinant provides a Lagrangian
\be
L=(\det h)^{1/2} d^2q =([\eta_{AB} z^A_1 z^B_1] [\eta_{AB} z^A_2
z^B_2]- [\eta_{AB} z^A_1 z^B_2]^2 )^{1/2} d^2q
\ee
on the jet manifold $J^1Z_Q$. This is the well known Nambu--Goto
Lagrangian of string theory \cite{hatf,polch}. It satisfies the
Noether identities (\ref{s60}).

\bigskip
\bigskip
\bigskip

\centerline{\large \bf IV. Quantum outcomes}

\section{Quantum master equation}

Discussing quantization of ACFT, we restrict our consideration to
the case of a vector bundle $Y\to X$ of classical field and ACFT
obeying  item (ii) of Theorem \ref{t6}, i.e., its BRST extension
$P^*_\infty\{N\}$ is characterized by a Lagrangian $L_E$
(\ref{w133}) and the BRST operator $u_E$ (\ref{w109}).  One can
quantize this BRST theory in the framework of perturbative QFT in
functional integral terms. This QFT is well formulated if a field
Lagrangian is non-degenerate. A problem is that the BRST extended
Lagrangian $L_E=\cL_E\om$ is necessarily degenerate. Indeed, it
obeys the classical master equation
\mar{f44}\beq
\{L_E,L_E\}=2\frac{\op\dl^\lto \cL_E}{\dl \ol z_a}\frac{\dl
\cL_E}{\dl z^a}\om =0 \label{f44}
\eeq
which is reducible Noether identities. To overcome this
difficulty, one often require that a BRST extended Lagrangian
 is a solution $L_h$ of the quantum master equation
\be
\{L_h,L_h\}=h\frac{\op\dl^\lto }{\dl \ol z_a}\frac{\dl }{\dl
z^a}\cL_h\om, \qquad h={\rm const.},
\ee
\cite{bat,fulp05,gom,fust}. Accordingly, a quantum BRST operator
is defined, and quantum BRST cohomology are studied.

\section{Gauge fixing procedure}

In order to make a Lagrangian $L_E$ non-degenerate, one can
replace antifields in $L_E$ with gauge-fixing terms
\cite{bat,fust,gom}. For this purpose, let us consider an odd
graded density $\Psi \om$ of antifield number 1 which depends on
original fields $s^A$ and ghosts $c^{r_k}$, $k=0,\ldots, N$, but
not antifields $\ol s_A$, $\ol c_{r_k}$, $k=0,\ldots, N$. In order
to satisfy these conditions, new field variables must be
introduced because all the ghosts are of negative antifield
numbers. Therefore, let us enlarge the BGDA $P^*\{N\}$ to the BGDA
$\ol P^*\{N\}$, possessing the basis
\be
 \{s^A, c^r, c^{r_1}, \ldots,
c^{r_N},c^*_r, c^*_{r_1}, \ldots, c^*_{r_N}, \ol s_A,\ol c_r, \ol
c_{r_1}, \ldots, \ol c_{r_N}\},
\ee
where $[c^*_{r_k}]=[c^{r_k}]$ and Ant$[c^*_{r_k}]=k+1$,
$k=0,\ldots,N$ \cite{ward}. Then one can choose $\Psi \om$ as an
element of $\ol P^{0,n}\{N\}$. It is traditionally called the
gauge-fixing fermion.

Let us replace all the antifields in the Lagrangian $L_E$
(\ref{w133}) with the gauge fixing terms
\be
\ol s_A=\frac{\dl\Psi}{\dl s^A}, \qquad \ol
c_{r_k}=\frac{\dl\Psi}{\dl c^{r_k}}, \qquad k=0,\ldots, N.
\ee
We obtain the Lagrangian
\mar{z13}\beq
L_\Psi= L + [u^A_E \frac{\dl \Psi}{\dl s^A} + \op\sum_{0\leq k\leq
N} u^{r_k}_E\frac{\dl \Psi}{\dl c^{r_k}}]\om =L + u_E(\Psi)\om
+d_H\si. \label{z13}
\eeq
A glance at the equalities
\be
u_E(L_\Psi)=u(L) + u_E(u_E(\Psi))\om +d_H\si =d_H\si'
\ee
shows that BRST operator $u_E$ (\ref{w109}) is a variational
symmetry of the Lagrangian $L_\Psi$. It however is not a gauge
symmetry of $L_\Psi$ if $L_\Psi$ depends on all the ghosts
$c^{r_k}$, $k=0,\ldots,N$, i.e., no ghost is a gauge parameter.
Therefore, we require that
\mar{z14}\beq
\frac{\dl \Psi}{\dl s^A}\neq 0, \qquad \frac{\dl \Psi}{\dl
c^{r_k}}\neq 0, \qquad k=0,\ldots, N-1. \label{z14}
\eeq
In this case, Noether identities for the Lagrangian $L_\Psi$
(\ref{z13}) come neither from the BRST symmetry $u_E$ nor the
equalities (\ref{f44}). One also put
\mar{z15}\beq
\Psi= \op\sum_{0\leq k\leq N} \Psi^{r_k}c^*_{r_k}. \label{z15}
\eeq
Finally, let $h^{r_k r'_k}$ be a non-degenerate bilinear form for
each $k=0,\ldots, N-1$ whose coefficients are either real numbers
or functions on $X$. Then a desired gauge-fixing Lagrangian is
written in the form
\mar{z16}\beq
L_{GF}=L_\Psi + \op\sum_{0\leq k\leq N}\frac12 h_{r_kr'_k}
\Psi^{r_k}\Psi^{r'_k}d^nx.  \label{z16}
\eeq

The BRST operator $u_E$ (\ref{w109}) fails to be a variational
symmetry of the Lagrangian (\ref{z16}), but it can be extended to
its variational symmetry
\be
\wh u=u_E- \op\sum_{0\leq k\leq N}\frac{\op\dr^\lto}{\dr
c^*_{r_k}}h_{r_kr'_k}\Psi^{r'_k},
\ee
though it is not nilpotent. Of course, the Lagrangian $L_{GF}$
and, accordingly,  the generating functional of perturbative QFT
essentially depends on a choice of the gauge-fixing fermion
$\Psi$. The generating functional is invariant under the
variations $\Psi+\dl\Psi$ of $\Psi$ if the gauge fixing Lagrangian
obeys the quantum master equation \cite{gom}.

\section{Green function identities}

In order to obtain the generating functional of BRST theory, one
replaces horizontal densities, depending on jets, with local
functionals (see item 12) evaluated for the jet prolongations of
sections  \cite{ala,barn98,barn,bran97,mccloud}. Note that such
functionals, in turn, define differential forms on functional
spaces \cite{castr3,ferr}. At the same time, there is the
following relation between the algebras of jets of classical
fields and the algebras of quantum fields \cite{ward2}.

Let us consider a Lagrangian field system on $X=\Bbb R^n$,
coordinated by $(x^\la)$. It is described by the GDA $P^*$
possessing a local basis
\mar{bb3}\beq
\{s^a, s^a_\la, s^a_{\la_1\la_2},\ldots,
s^a_{\la_1\ldots\la_k},\ldots\}. \label{bb3}
\eeq
Let $L\in P^{0,n}$ be a non degenerate Lagrangian. Let us quantize
this Lagrangian system in the framework of perturbative Euclidean
QFT. We suppose that $L$ is a Lagrangian of Euclidean fields on
$X=\Bbb R^n$. The key point is that the algebra of Euclidean
quantum fields $B_\Phi$ as like as $P^0$ is graded commutative
\cite{sard91,sard02a,ward2}. It is generated by elements
$\f^a_{x\La}$, $x\in X$. For any $x\in X$, there is a homomorphism
\mar{z45}\beq
\g_x: f_{a_1\ldots a_r}^{\La_1\ldots\La_r} s^{a_1}_{\La_1}\cdots
s^{a_r}_{\La_r} \mapsto f_{a_1\ldots a_r}^{\La_1\ldots\La_r}(x)
\f^{a_1}_{x\La_1}\cdots \f_{x\La_r}^{a_r}, \qquad f_{a_1\ldots
a_r}^{\La_1\ldots\La_r}\in C^\infty(X), \label{z45}
\eeq
of the algebra $P^0$ of classical fields to the algebra $B_\Phi$
which sends the basis elements $s^a_\La\in P^0$ to the elements
$\f^a_{x\La}\in B_\Phi$, and replaces coefficient functions $f$ of
elements of $\cP^0$ with their values $f(x)$ at a point $x$. Then
a state $\lng.\rng$ of $B_\Phi$ is given by symbolic functional
integrals
\mar{z49}\ben
&& \lng\f^{a_1}_{x_1}\cdots \f^{a_k}_{x_k}\rng=\frac{1}{\cN}\int
\f^{a_1}_{x_1}\cdots \f^{a_k}_{x_k} \exp\{-\int
\cL(\f^a_{x\La})d^nx\}\op\prod_x [d\phi_x^a], \label{z49}\\
&& \cN=\int \exp\{-\int \cL(\f^a_{x\La})d^nx\}\op\prod_x
[d\phi_x^a], \nonumber\\
&& \cL(\f^a_{x\La})=\cL(x,\g_x(s^a_\La)), \nonumber
\een
which restart complete Euclidean Green functions in the Feynman
diagram technique.

Due to homomorphisms (\ref{z45}), any graded derivation $\vt$  of
$P^0$ induces the graded derivation
\be
 \wh \vt: \f^a_{x\La}\to
(x, s^a_\La)\to u^a_\La(x,s^b_\Si)\to u^a_\La(x,\g_x(s^b_\Si))=\wh
\vt^a_{x\La}(\f^b_{x\Si})
\ee
of the algebra of quantum fields $B_\Phi$ \cite{ward2}. With an
odd parameter $\al$, let us consider the automorphism
\be
\wh U=\exp\{\al \wh \vt\}=\id +\al\wh \vt
\ee
of the algebra $B_\Phi$. This automorphism yields a new state
$\lng.\rng'$ of $B_\Phi$ given by the equalities
\mar{bb10}\ben
&& \lng \f^{a_1}_{x_1}\cdots \f^{a_k}_{x_k}\rng= \lng \wh
U(\f^{a_1}_{x_1})\cdots \wh U(\f^{a_k}_{x_k})\rng'=  \label{bb10}\\
&& \qquad \frac{1}{\cN'}\int \wh U(\f^{a_1}_{x_1})\cdots \wh
U(\f^{a_k}_{x_k}) \exp\{-\int \cL(\wh
U(\f^a_{x\La}))d^nx\}\op\prod_x [d\wh
U(\phi_x^a)], \nonumber \\
&& \cN'=\int \exp\{-\int \cL(\wh U(\f^a_{x\La}))d^nx\}\op\prod_x
[d\wh U(\phi_x^a)]. \nonumber
\een
It follows from the first variational formula (\ref{g107}) that
\be
\int \cL(\wh U(\f^a_{x\La}))d^nx =\int (\cL(\f^a_{x\La}) + \al \wh
\vt_x^a\cE_{xa})\om,
\ee
where $\cE_{xa}=\g_x(\cE_a)$ are the variational derivatives. It
is a property of symbolic functional integrals that
\be
\op\prod_x[d\wh U(\phi_x^a)]=(1+\al\int \frac{\dr \wh \vt^a_x}{\dr
\f^a_x}d^nx)\op\prod_x[d\phi_x^a]=(1+\al {\rm Sp}(\wh \vt))
\op\prod_x[d\phi_x^a].
\ee
 Then the
equalities (\ref{bb10}) result in the identities
\mar{z62'}\ben
&& \lng\wh \vt(\f^{a_1}_{x_1}\cdots \f^{a_k}_{x_k})\rng +
\lng\f^{a_1}_{x_1}\cdots \f^{a_k}_{x_k}({\rm Sp}(\wh \vt) -\int
\wh
\vt_x^a\cE_{xa}d^nx)\rng - \label{z62'}\\
&& \qquad \lng\f^{a_1}_{x_1}\cdots \f^{a_k}_{x_k}\rng\lng {\rm
Sp}(\wh \vt) -\int \wh \vt_x^a\cE_{xa}d^nx\rng =0. \nonumber
\een
for complete Euclidean Green functions (\ref{z49}).

In particular, if $\vt$ is a variational symmetry of a Lagrangian
$L$, the identities (\ref{z62'}) are the Ward identities
\mar{bb11}\beq
\lng\wh \vt(\f^{a_1}_{x_1}\cdots \f^{a_k}_{x_k})\rng +
\lng\f^{a_1}_{x_1}\cdots \f^{a_k}_{x_k}{\rm Sp}(\wh \vt)\rng -
\lng\f^{a_1}_{x_1}\cdots \f^{a_k}_{x_k}\rng\lng {\rm Sp}(\wh
\vt)\rng =0, \label{bb11}
\eeq
generalizing the Ward (Slavnov--Taylor) identities in gauge theory
\cite{bran98,fust,mccloud}.

If $\vt=c^a\dr_a$, $c^a=$const, the identities (\ref{z62'}) take
the form
\mar{bb13}\ben
&& \op\sum_{r=1}^k (-1)^{[a]([a_1]+\cdots+[a_{r-1}])}
\lng\f^{a_1}_{x_1}\cdots \f^{a_{r-1}}_{x_{r-1}} \dl^{a_r}_a
\f^{a_{r+1}}_{x_{r+1}}\cdots
\f^{a_k}_{x_k})\rng -  \label{bb13}\\
&& \qquad \lng\f^{a_1}_{x_1}\cdots \f^{a_k}_{x_k}(\int
\cE_{xa}d^nx)\rng + \lng\f^{a_1}_{x_1}\cdots
\f^{a_k}_{x_k}\rng\lng \int \wh \cE_{xa}d^nx\rng =0. \nonumber
\een
One can think of them as being equations for complete Euclidean
Green functions. Clearly, the expressions (\ref{z62'}) --
(\ref{bb13}) are singular, unless one follows regularization and
renormalization procedures, which however can induce additional
anomaly terms.

\end{document}